\documentclass{article}
\usepackage{jcappub}

\def\lbldef#1#2{\expandafter\gdef\csname #1\endcsname {#2}}

\def\href#1#2{#2}

\usepackage{graphicx}
\usepackage{epsfig}
\usepackage{amsmath}
\usepackage{amssymb}

\title{New constraints on cosmological parameters and neutrino properties
using the expansion rate of the Universe to $z\sim1.75$}

\author[1]{Michele Moresco}
\author[2]{, Licia Verde}
\author[3]{, Lucia Pozzetti}
\author[2]{, Raul Jimenez}
\author[1]{and Andrea Cimatti}

\affiliation[1]{Dipartimento di Astronomia, Universit\'a degli Studi di Bologna, via Ranzani 1, I-40127, Bologna, Italy}
\affiliation[2]{ICREA \& Institute of Sciences of the Cosmos (ICC), Universitat de Barcelona (IEEC-UB), Mart' i Franqus, 1, Barcelona 08028, Spain}
\affiliation[3]{INAF - Osservatorio Astronomico di Bologna, via Ranzani 1, 40127 Bologna, Italy}

\emailAdd{michele.moresco@unibo.it, licia.verde@icc.ub.edu, lucia.pozzetti@oabo.inaf.it, raul.jimenez@icc.ub.edu, a.cimatti@unibo.it}

\abstract{
We have assembled a compilation of observational Hubble parameter measurements estimated with the differential evolution of cosmic chronometers, 
in the redshift range $0<z<1.75$. This sample has been used, in combination with CMB data and with the most recent estimate of the Hubble constant $H_{0}$, 
to derive new constraints on several cosmological parameters. 
The new Hubble parameter data are very useful to break some of the parameter degeneracies present in CMB-only analysis, and to constrain possible 
deviations from the standard (minimal) flat $\Lambda$CDM model. The $H(z)$ data are especially valuable in constraining $\Omega_k$ and $\Omega_{\rm DE}$ 
in models that allow a variation of those parameters, yielding constraints that are competitive with those obtained using Supernovae and/or baryon acoustic oscillations.
We also find that our $H(z)$ data are important to constrain parameters that do no affect directly the expansion history, by breaking or reducing degeneracies with 
other parameters. We find that $N_{\rm rel}=3.45\pm0.33$ using WMAP 7-years data in combination with South Pole Telescope data and our $H(z)$ determinations 
($N_{\rm rel}=3.71\pm0.45$ using Atacama Cosmology Telescope data instead of South Pole Telescope). We exclude $N_{\rm rel}>4$ at 95\% CL 
(74\% CL) using the same datasets combinations. We also put competitive limits on the sum of neutrino masses, $\Sigma m_{\nu}<0.24$ 
eV at 68\% confidence level. These results have been proven to be extremely robust to many possible systematic effects, such as the initial choice of stellar population 
synthesis model adopted to estimate $H(z)$ and the progenitor-bias.
}

\begin{document}

\maketitle

%%%%%%%%%%%%%%%%%%%%%%%%%%%%%%%%%%%%%%%%%%%%%%%%%%%%%%%%%%%%%%%%%%%%%%%%%%%%%%%%%%%%%%%%%%%

\section{Introduction}

The accelerated expansion of the Universe \cite{Riess1998,Perlmutter1999} has become, during the last decades,
a stronghold of modern cosmology, witnessed by a lot of independent probes. 
However the real physical mechanism driving 
such accelerated expansion is still unknown; one possible explanation is that it could be due to an unknown 
form of energy, ``dark energy'', which in its simplest embodiment is a cosmological constant, but other possibilities could include
a breakdown of general relativity on large scales or an effect of interpreting the observation using a metric 
which is not correct for our inhomogeneous Universe.

At present, the most economical framework to describe all available observations is the $\Lambda$CDM paradigm, 
in which the Universe is spatially flat, its energy budget is dominated by a cosmological constant, and 
there are three massless neutrinos. The $\Lambda$CDM model is able to describe the evolution of the 
Universe with a minimal number of cosmological parameters; current data constrain these parameters at the \% level.
However, this is just a phenomenological description, as we do not have 
full physical understanding of many of these parameters (despite their small errorbars).

One approach to advance on this issue is to constrain cosmological parameters that describe deviations from this ``minimal'' model.
Several different probes may be -- and have been -- used to set constraints on cosmological 
parameters, and especially on dark energy parameters, each one having its strengths and weaknesses; among them, the most well-known 
are the Cosmic Microwave Background (hereafter CMB), the Baryonic Acoustic Oscillation (hereafter BAO), 
the Supernovae type Ia (hereafter SNe), plus probes of the growth of structure via weak lensing studies and cluster 
of galaxies abundance. 
%(the literature on the subject is enormous: ads returns over $20,000$ hits with the words ``cosmological parameters" in the abstract). 
A good reference that gives an overview of combining different probes to obtain cosmological parameters is the series of papers by the WMAP team (e.g., \cite{Verde2003,Komatsu2011}). 

The above probes constrain the expansion history using geometric measurements: e.g., BAO use {\it standard rulers} and SNe {\it standard(-izable) candles}. 
A complementary technique to track directly the expansion history of the Universe is to use
massive and passively evolving early-type galaxies (hereafter ETGs) as ``cosmic chronometers'' \cite{Jimenez2002}, 
thus providing {\it standard(-izable) clocks} in the Universe.
The basic idea underlying this approach is based on the measurement
of the differential age evolution as a function of redshift of these chronometers, which provides a direct estimate
of the Hubble parameter $H(z)=-1/(1+z)dz/dt \simeq -1/(1+z)\Delta z/\Delta t$.
The main strength of this approach is the reliance on the measurement of a differential quantity, 
$\Delta z/\Delta t$, which provides many advantages in minimizing many common issues 
and systematic effects (for a detailed discussion, see \cite{Moresco2012}). Moreover, compared with other techniques, 
this approach provides a direct measurement of the Hubble parameter, and not of its integral, as in the case of e.g., 
SNe or angular/angle-averaged BAO. For this reason the Hubble parameter estimated with this approach 
has been referred as ``observational Hubble parameter data'' (OHD, see e.g. \cite{Zhang2010,Ma2011}).

There has been significant progress both in the theoretical understanding of the method and control 
of systematics uncertainties \cite{Moresco2011,Moresco2012} and in the improvement of observational data \cite{Moresco2012}. 
This paper is the second in a series of three \cite{Moresco2012,Jimenez2012} where we present 
new $H(z)$ determinations \cite{Moresco2012} and interpret their result in terms of cosmological parameters constraints (this paper) 
and a model independent reconstruction of the dark energy potential \cite{Jimenez2012}. 

The paper is organized as follows. In Sect. \ref{sec:theory} we explore the sensitivity of the Hubble parameter
to the various cosmological parameters and its statistical power in constraining them. The data sample
comprising all the observational $H(z)$ measurements used in this analysis is provided in Sect. \ref{sec:data};
in this section we also describe the additional datasets used to obtain constraints on the cosmological parameters,
i.e. the cosmic microwave background measurements from WMAP (7 years results) \cite{Larson2011}, the results obtained from
baryonic acoustic oscillations analysis \cite{Percival2010}, and finally the ones obtained from Supernovae type Ia \cite{Hicken2009}.
In Sect. \ref{sec:results} we present our results in constraining the deviations from the standard $\Lambda$CDM model,
studying both cases in which the expansion history of the Universe helps in constraining those parameters, and
cases in which the constraints are set just by breaking some degeneracies between the parameters. 
We also discuss the reliability of our constraints against the choice of stellar population synthesis model adopted to estimate the
Hubble parameter, and estimate the effect of the progenitor bias. Finally, in Sect. \ref{sec:future} we will explore how the constraints on the cosmological parameters will be
improved with more accurate measurements of $H(z)$ over a wider redshift range, which future surveys will make soon available.

\section{Theoretical background}
\label{sec:theory}

\begin{figure}[t!]
\begin{center}
\includegraphics[angle=-90, width=0.75\textwidth]{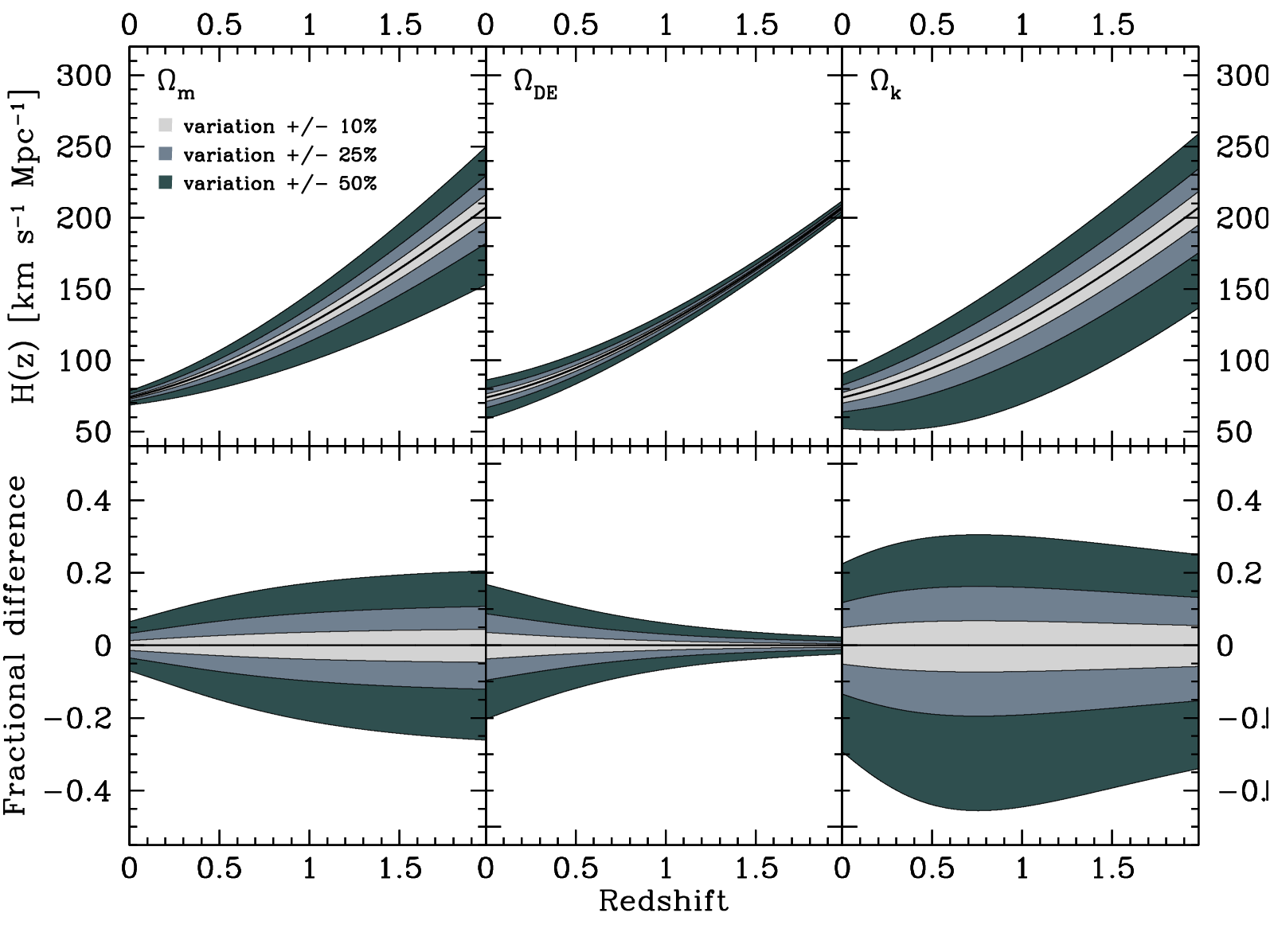}
\includegraphics[angle=-90, width=0.75\textwidth]{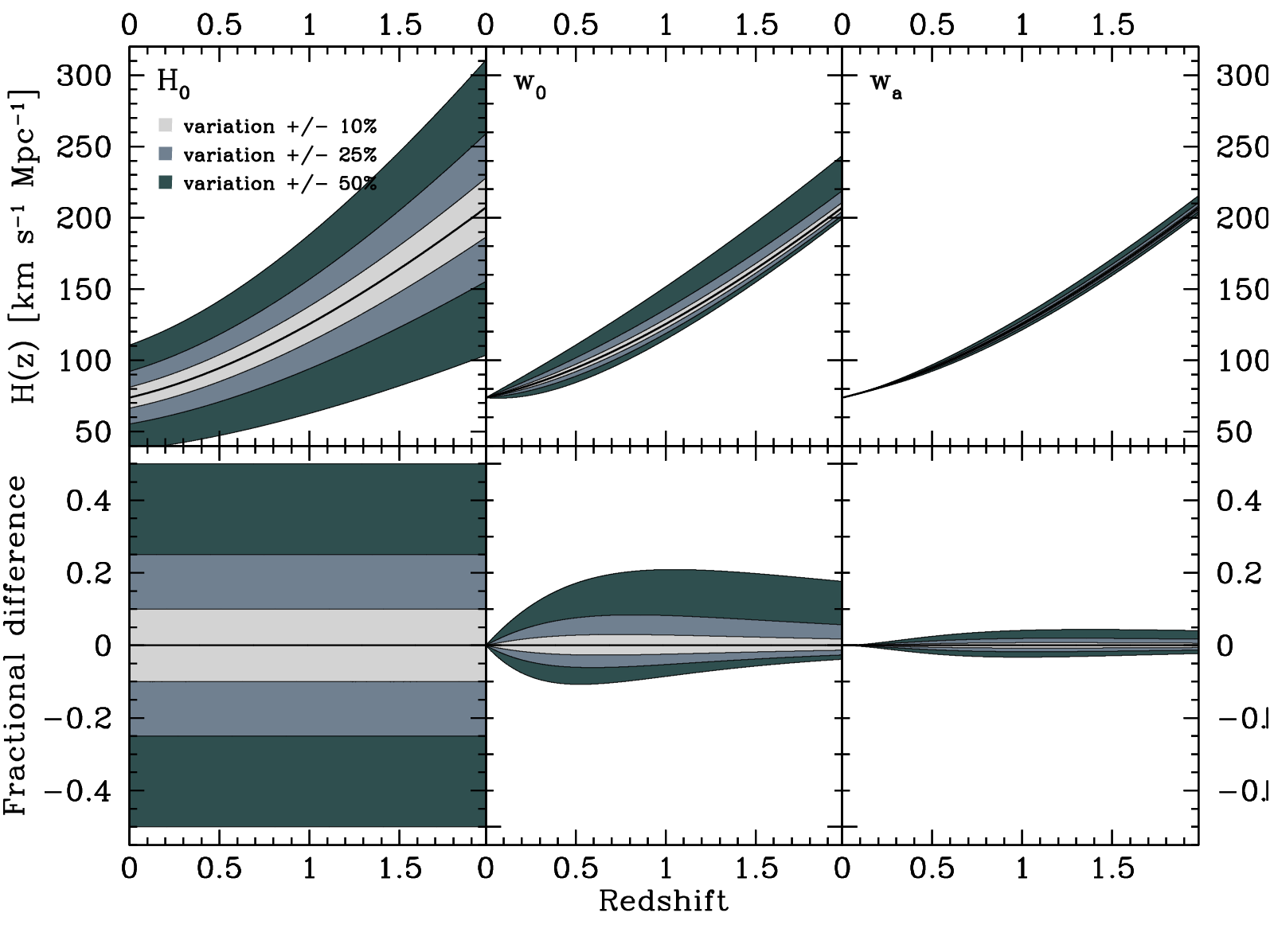}
\caption{The dependence of the Hubble parameter $H(z)$ on the cosmological parameters. One cosmological parameter at the time is being changed with respect to the fiducial model by 10, 25 and 50\%. In each of the two panels, the upper (lower) rows show the absolute (relative) effect of this variation on the Hubble parameter as a function of redshift.
\label{fig:Hztheor}}
\end{center}
\end{figure}

The detailed formalism at the base of the {\it cosmic chronometers} approach is introduced in Ref.~\cite{Jimenez2002},
and explained in detail in e.g. Ref.~\cite{Moresco2012}, considering also all the possible sources of uncertainty and related issues; we refer to those references for a comprehensive discussion.

In this section we analyze, from the theoretical point of view, the statistical power in constraining several cosmological parameters of the measurements of the Hubble parameter $H(z)$. This section provides physical understanding of how the change of different cosmological parameters affects $H(z)$, which will then be useful to understand the degeneracies among cosmological parameters we can hope to break with the addition of accurate measurements of the expansion history of the universe to, e.g., CMB constraints.

Assuming a Friedmann-$\mathrm{Lema\hat{\i}tre}$-Robertson-Walker (FLRW) metric, it is possible to write the Hubble parameter
as a function of the cosmological parameters (including curvature and dark energy with redshift-dependent equation of state parameter) as follows:
\begin{equation}
H(z)=H_{0}\left\{\Omega_{m}(1+z)^{3}+\Omega_{k}(1+z)^{2}+\Omega_{\rm DE}(1+z)^{3(1+w_{0}+w_{a})}e^{-3w_{a}\frac{z}{1+z}}\right\}^{1/2}
\label{eq:Hztheor}
\end{equation}
where $H_{0}$ is the Hubble constant, $\Omega_{i}$ is the adimensional energy density parameter for each specie in the Universe (i.e. matter $\Omega_{m}$, curvature $\Omega_{k}$ and dark energy $\Omega_{\rm DE}$; we have neglected radiation since its contribution is irrelevant in the late Universe), and dark energy Equation of State (EoS) parameter
as suggested in Ref.~\cite{Chevallier2001,Linder2003}, $w(z)=w_{0}+w_{a}(z/(1+z))$.

To illustrate the dependence of the observed $H(z)$ on cosmological parameters, we fix a fiducial cosmology,
taken from the latest $H_{0}$ estimate \cite{Riess2011} and from the analysis of the WMAP 7-years data \cite{Komatsu2011}
(i.e. $H_{0}=73.8\mathrm{km\,s^{-1}Mpc^{-1}}$, $\Omega_{k,0}=0$, $\Omega_{m,0}=0.27$, $\Omega_{\mathrm{DE},0}=0.73$, $w_{0}=-1$, $w_{a}=0$), and 
perturb it by varying one parameter at a time by $\pm$10\%, 25\%, 50\% (in the case of $\Omega_{k}$ and $w_{a}$, we explored 
the values 0.1, 0.25 and 0.5). We then estimate the absolute and fractional difference on $H(z)$,
$(H(z)_{\mathrm{ref}}-H(z)_{\mathrm{pert}})/H(z)_{\mathrm{ref}}$. The results are shown in Fig. \ref{fig:Hztheor}:
the upper plot shows the effect on $H(z)$ of the adimensional density parameters $\Omega_{m}$,
$\Omega_{\mathrm{DE}}$ and $\Omega_{k}$, showing respectively the Hubble parameter (plots on upper rows)
and the fractional difference (plots on lower rows). In the lower plot we show, as in the upper plot, the dependence
on the Hubble constant, dark energy equation-of-state parameter $w_{0}$ and $w_{a}$. 

As it is evident from Eq. \ref{eq:Hztheor}, the dependence on the Hubble constant $H_{0}$ is linear;
what is less intuitive from the form of the equation is the detailed dependence on the other parameters.
We found that the Hubble parameter is particularly sensitive to the curvature parameter, showing a percentage variation of
$\sim30-40$\% in $H(z)$ with $\Omega_{k}=0.5$; the dependence on the other density parameter is
less pronounced and redshift dependent, with a percentage variation up to 25\% at $z\sim2$
with a variation of 50\% in $\Omega_{m}$, and a percentage variation up to 25\% at $z\sim0$
with a variation of 50\% in $\Omega_{\mathrm{DE}}$. The less trivial parameters to be constrained are,
as already known, the dark energy equation-of-state parameters $w_{0}$ and $w_{a}$: for the first one,
a variation of 50\% produces a non-symmetric percentage variation, that could go up to 20\% when moving towards less negative values of $w_{0}$,
but being less sensitive to more negative values of $w_{0}$. The parameter $w_{a}$ is clearly the most
elusive one, showing a percentage variation of only 5\% corresponding to the maximum value explored, i.e. $w_{a}=0.5$.

\section{Observational Data \& Method}
\label{sec:data}
\begin{figure}[t!]
\begin{center}
\includegraphics[angle=-90, width=0.95\textwidth]{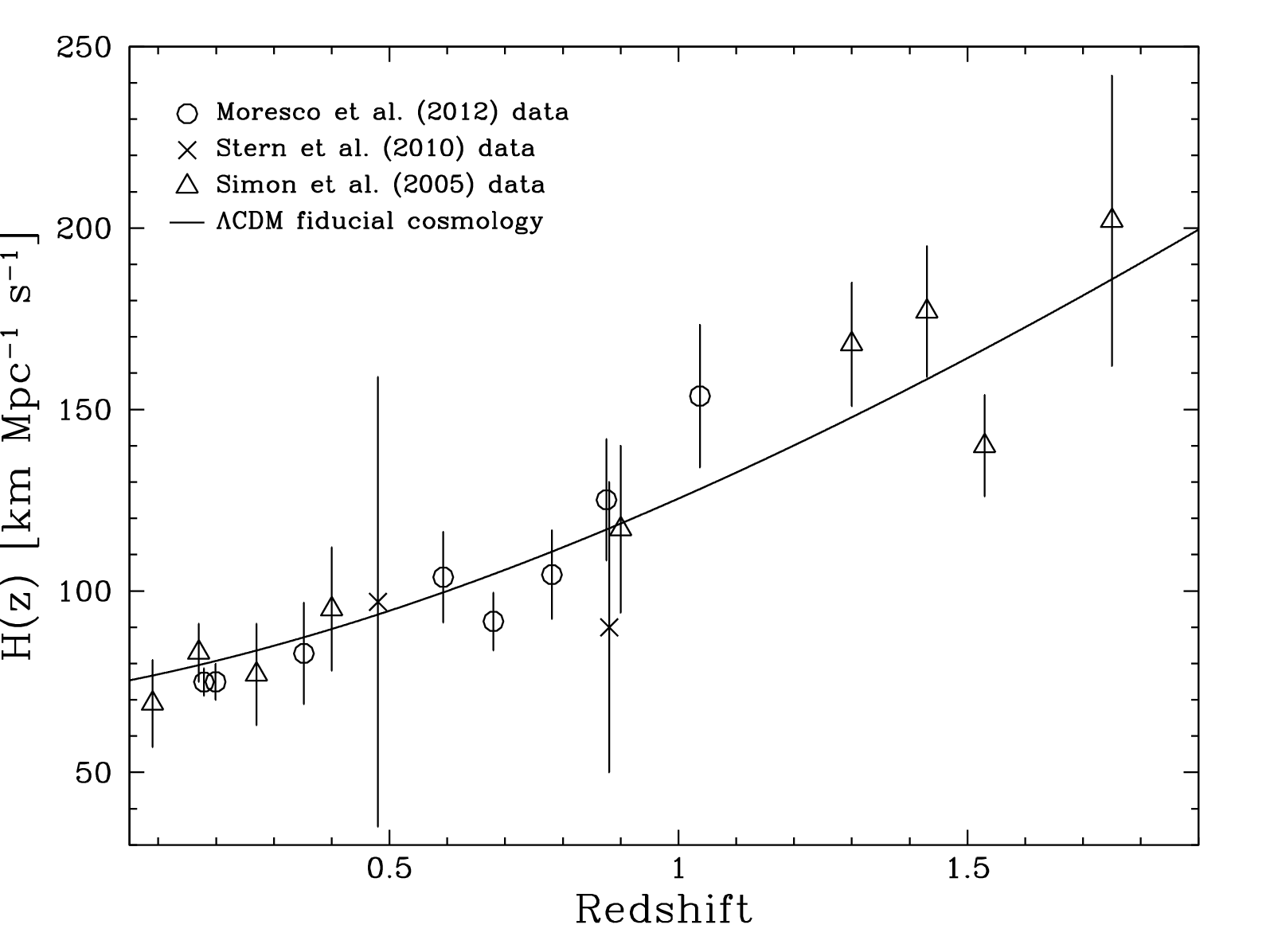}
\caption{Hubble parameter measurements. Open circle points correspond to $H(z)$ estimates from Ref.~\cite{Moresco2012}, open triangle points from Ref.~\cite{Simon2005}, and starred points are taken from Ref.~\cite{Stern2010}. We also show a fiducial $\Lambda$CDM cosmology, with $H_{0}=73.8$, $\Omega_{k,0}=0$, $\Omega_{m,0}=0.27$, $\Omega_{\mathrm{DE},0}=0.73$, $w_{0}=-1$, $w_{a}=0$. %For a comparison, we show also the $H(z)$ estimates obtained with different methods, obtained from the analysis of SNe and radio galaxies \cite{Daly2008} and from the two-dimensional two point correlation function from SDSS-DR7 LRGs \cite{Chuang2011}.
\label{fig:Hzdata}}
\end{center}
\end{figure}
We combined the observational constraints on the Hubble parameter available from the literature to obtain the widest redshift lever arm.
%We decided to use all the available data in literature that has obtained a measure of the observational Hubble parameter, and to combine them to obtain the widest and most homogeneous sampling possible. 

We started from the dataset provided by Ref.~\cite{Simon2005}, which obtained 9 determinations of the Hubble 
parameter in the range $0.1<z<1.75$ from the relative dating of 32 passively evolving galaxies.
Ref.~\cite{Stern2010} obtained high-quality spectra of red-envelope galaxies in 24 galaxy 
clusters in the redshift range $0.2<z<1.0$, yielding 2 more observations of the Hubble parameter
at $z\sim0.5$ and $z\sim0.9$. Finally, we added the latest 8 new high-accuracy estimates 
of $H(z)$ provided by Ref.~\cite{Moresco2012}. They analyzed the spectroscopic evolution of
a large sample of 11324 early type galaxies, extracted from several spectroscopic surveys to ensure a high statistic and uniform coverage in the redshift range $0.15<z<1.1$.

There are many systematic sources of uncertainty that can bias this kind of analysis; they can all be understood in terms 
of degeneracies between parameters which can in some way mimic an age-effect
(e.g. the stellar metallicity). In the following we list them, and explain how they have been treated 
and their effect propagated in the total error budget of the $H(z)$ measurements.
We refer to the original papers \cite{Simon2005,Stern2010,Moresco2012} for the detailed discussion
of how they have been taken into account.
\begin{itemize}
\item {\bf Stellar Metallicity and Star Formation History.} It is well known that the effect of a change in stellar metallicity ($Z$)
or in Star Formation History ($SFH$) is degenerate with the effect of a change in age, so that 
the multi-band spectral energy distribution of a galaxy with a given age and metallicity may be very similar to that of a galaxy with 
lower age and higher metallicity. Analogous degeneracies are present when analyzing galaxy spectra, especially if
wavelengths below 4000~\AA~are not observed (see \cite{Stern2010}).
All the samples considered take into account this systematic effect in the total $H(z)$ error.
In Ref.~\cite{Simon2005}, the sample has been selected by choosing ETGs with spectra best-fitted by a passive 
population, and their ages have been estimated marginalizing over the metallicity and star formation history. 
In Ref.~\cite{Stern2010} to estimate the best age a $\chi^{2}$ fit to the spectra has been performed using a large 
grid of ages and metallicities, and the errorbars have been evaluated also in this case by marginalizing over the metallicity;
the star formation has been assumed to occur in a single burst of short duration, and galaxies with poor values
of $\chi^{2}$ have been discarded. In Ref.~\cite{Moresco2012} the ``cosmic chronometers'' approach has been developed to be
even less dependent on all the degeneracies previously discussed. The ETGs sample has been selected to represent the most 
massive ($log_{10}(\mathcal{M/M_{\odot}})\gtrsim11$) and passive galaxies. Instead of using the age estimated from the best-fit to the
spectral energy distribution or to the spectrum, the 4000~\AA~break (D4000) has been studied, which is a direct observable present in ETGs spectra 
known to correlate with age (and metallicity) of the galaxy. To calibrate this index, two different theoretical stellar population synthesis 
models have been studied. Finally, since in Ref.~\cite{Moresco2011} it has been noted the existence of a clear dependence of the D4000-z relation 
on the stellar mass, two different mass subsamples have been obtained, and the Hubble parameter obtained separately from the two. 
Since the obtained results show a good agreement, the $H(z)$ measurements obtained from the two mass-subsamples have been averaged.
The total errorbar of the Hubble parameter has been estimated by considering both the statistical errors and the systematic
uncertainty due to star formation history and metallicity.
\item {\bf Stellar Population Synthesis Models.} Also the choice of different Stellar Population Synthesis (SPS) models, used to estimate the age 
or to calibrate the D4000-age relation, may bias the estimate of $H(z)$, and hence be a possible source of systematic uncertainty. Ref.~\cite{Stern2010}
explored the sensitivity of the age estimate obtained by fitting galaxy spectra, finding a general agreement between the measurements performed
using different models in particular when is available the wavelength coverage below 4000~\AA~. In the work of 
Ref.~\cite{Moresco2012} it is discussed in detail the effect of considering different stellar population synthesis models, by studying both Bruzual 
\& Charlot (2003) (\cite{Bruzual2003}, hereafter BC03) and Maraston \& Str{\"o}mb{\"a}ck (2011) (\cite{Maraston2011}, hereafter MaStro) models. 
Since all the other previous analysis in literature provided $H(z)$ measurements only for BC03 models, for the sake of uniformity, in this work we decided to 
combine estimates obtained with BC03 models. In Sect. \ref{sec:SPSmodels} we discuss the reliability of our results by analyzing also the constraints 
obtained with the MaStro models from Ref.~\cite{Moresco2012}, to estimate the sensitivity of our results to different stellar population synthesis models.
\item {\bf Progenitor Bias} The progenitor bias (e.g. see \cite{vanDokkum1996}) refers to the effect for which high-redshift samples of ETGs
might not be statistically equivalent to the low-redshift samples, but a subset of the low-redshift samples biased toward higher ages.
The measurements of Ref.~\cite{Simon2005} and Ref.~\cite{Stern2010} are based on the estimate of the upper and redder envelope
of the age-z relations, probing the oldest galaxies at each redshift. Hence they are by definition not affected by such a problem. On the contrary,
Ref.~\cite{Moresco2012} estimates $H(z)$ from the median $D4000-z$ relation of a mass-selected sample, and hence may be biased; 
in Sect. \ref{sec:progbias} we quantify this effect, and its influence on the total errors.
\end{itemize}

\begin{table}[h!]
\begin{center}
\begin{tabular}{lllcl}
\hline \hline
$z$ & $H(z)$ & $\sigma_{H(z)}$ & Ref.\\
\hline
0.090 & 69 & 12 & \cite{Simon2005}\\
0.170 & 83 & 8 & \cite{Simon2005}\\
0.179 & 75 & 4 & \cite{Moresco2012}\\
0.199 & 75 & 5 & \cite{Moresco2012}\\
0.270 & 77 & 14 & \cite{Simon2005}\\
0.352 & 83 & 14 & \cite{Moresco2012}\\
0.400 & 95 & 17 & \cite{Simon2005}\\
0.480 & 97 & 62 & \cite{Stern2010}\\
0.593 & 104 & 13 & \cite{Moresco2012}\\
0.680 & 92 & 8 & \cite{Moresco2012}\\
0.781 & 105 & 12 & \cite{Moresco2012}\\
0.875 & 125 & 17 & \cite{Moresco2012}\\
0.880 & 90 & 40 & \cite{Stern2010}\\
1.037 & 154 & 20 & \cite{Moresco2012}\\
1.300 & 168 & 17 & \cite{Simon2005}\\
1.430 & 177 & 18 & \cite{Simon2005}\\
1.530 & 140 & 14 & \cite{Simon2005}\\
1.750 & 202 & 40 & \cite{Simon2005}\\
\hline \hline
\end{tabular}
\caption{$H(z)$ measurements (in units [$\mathrm{km\,s^{-1}Mpc^{-1}}$]) and their errors. This dataset can be downloaded at the address http://www.astronomia.unibo.it/Astronomia/Ricerca/Progetti+e+attivita/cosmic\_chronometers.htm (alternatively http://start.at/cosmicchronometers).}
\label{tab:HzBC03}
\end{center}
\end{table}

In summary, we built a sample of 19 observational $H(z)$ measurements in the range $0.1<z<1.75$, spanning almost 10 Gyr of cosmic time; these values 
are reported in Tab. \ref{tab:HzBC03} and shown in Fig. \ref{fig:Hzdata}. This dataset represents the widest sample of observational constraints on the Hubble parameter available.
%For a comparison, we have also plotted the $H(z)$ measurements obtained with completely different approach: Ref.~\cite{Daly2008} obtained estimates of the Hubble parameter
%from the analysis of 192 Sne and 30 radio galaxies, while Ref.~\cite{Chuang2011} get informations about $H(z=0.35)$ by studying the two-dimensional two point correlation 
%function of the Luminous Red Galaxies in the SDSS Data Release 7 catalog (SDSS-DR7 LRGs). As it is evident from the figure, the estimates obtained with different techniques
%are in good agreement, and the advantage of the OHD is to obtain high accuracy measurements over a wide redshift range.

To set constraints on cosmological parameters, we decided to combine the OHD just described with the latest results from the WMAP 7-year release 
(hereafter WMAP7yr) \cite{Larson2011} and with South Pole Telescope \cite{Keisler2011} and Atacama Cosmology Telescope \cite{Dunkley2011,Das2011} 
(alternatively) measurements of the CMB temperature power spectrum damping tail. To complete the OHD, we will also use the latest and most precise 
estimate of the Hubble constant $H_{0}$ \cite{Riess2011}, which provides the local value of the Hubble parameter. Our aim, as stated before, is to 
constrain cosmological parameters using only CMB temperature data and OHD, thus combining only two probes. For a comparison, we will also show 
constraints from the combination of CMB \cite{Larson2011}, BAO \cite{Percival2010} and SNe data \cite{Hicken2009}. 
To estimate the relative importance of the $H(z)$ measurements and of the Hubble constant, we will also analyze the constraints obtained by combining to the CMB data
individually the Hubble constant $H_{0}$ and the OHD.
%When needed, we augmented the WMAP7yr data with small scale measurements of CMB temperature from ground-based experiments.

Monte Carlo Markov (MCMC) chains for the WMAP 7yr data have been obtained from the LAMBDA web 
site\footnote{www.lambda.gsfc.gov} and they have been importance-sampled with the $H(z)$ and $H_{0}$ data.
We considered different cosmologies, to see how much the use of the OHD can help in improving the constraints on different cosmological parameters.
The only case not covered by the LAMBDA web site is the open $w(z)$CDM model, i.e. where the dark-energy equation of state is parameterized as 
$w(z)=w_{0}+w_{a}(z/(1+z))$); we decided, therefore, to study the MCMC run specifically for this case by Ref.~\cite{Stern2010}, to explore
how much the new $H(z)$ data help in constraining specifically the dark energy parameters. Those chains, built allowing for arbitrary curvature and for
an evolving dark energy equation of state parametrized as described above, run on WMAP 5yr data and assume a constraint on $H_{0}$ from the Hubble 
Key Project \cite{Freedman2001}.

\section{Results}
\label{sec:results}
For a simple flat $\Lambda$CDM, the addition of OHD measurements does not change appreciably the parameters constraints: within this model the parameters 
are tightly constrained by CMB data alone. Still we note that OHD and CMB data are nicely consistent.
As anticipated, OHD measurements are however extremely useful in breaking CMB parameters degeneracies for models beyond the ``minimal'' $\Lambda$CDM model.
In particular we will show their fundamental importance in models which allow a time-variability of the dark energy EoS, where despite its extreme accuracy, the
$H_{0}$ alone is not able to set precise constraints to the cosmological parameters.
We therefore explore two different scenarios. In Sect. \ref{sec:resdirect} we show how $H(z)$ measurements can set constraints on the parameters
which can influence the expansion history of the Universe, such as $\Omega_{m}$, $\Omega_{\rm DE}$, $\Omega_{k}$, $w_{0}$, and $w_{a}$; in Sect. \ref{sec:resindirect} 
we show the importance of the OHD in breaking degeneracies between parameters that have an effect on H(z) and parameters that do not. This allow us to use our $H(z)$ 
data to constrain parameters which do not have a direct effect of $H(z)$ (but that are somewhat degenerate with the expansion history), such as the number of the 
relativistic species in the Universe $N_{\rm rel}$ and the sum of the neutrino masses $\Sigma m_{\nu}$.

\subsection{Constraints on deviations from $\Lambda$CDM that affect directly the expansion history}
\label{sec:resdirect}
\begin{figure}[t!]
\begin{center}
\includegraphics[angle=0, width=0.85\textwidth]{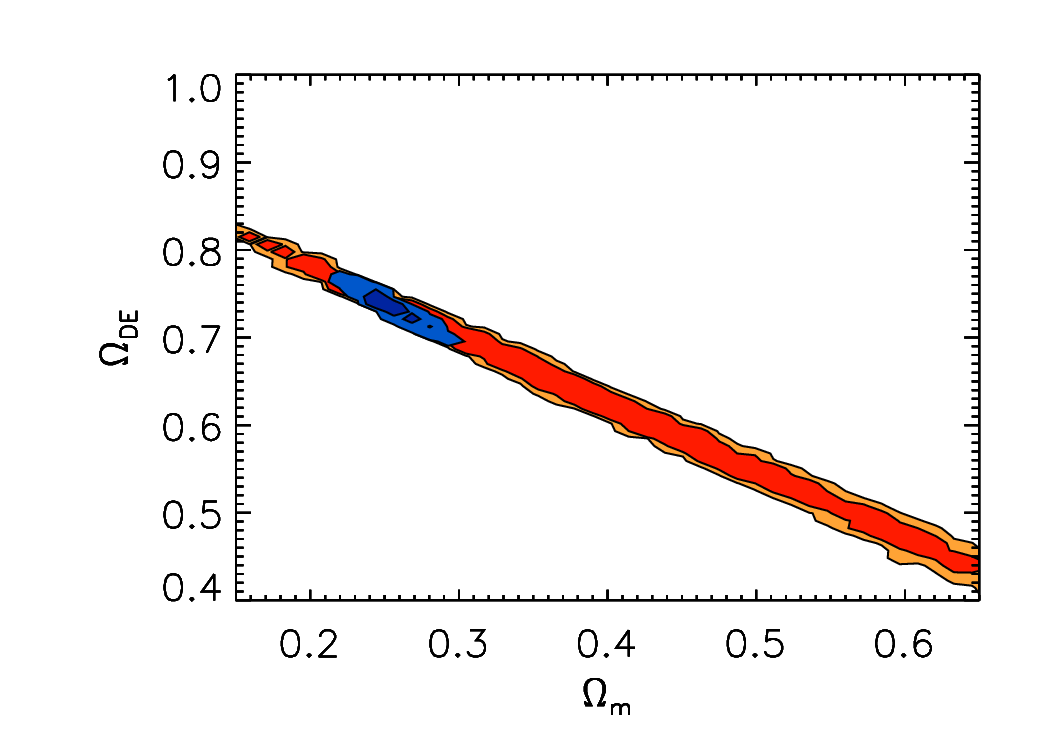}
\end{center}
\caption{1 and 2 sigma joint constraints in the $\Omega_{m}-\Omega_{\mathrm{DE}}$ plane
obtained for an open-$\Lambda$CDM model with WMAP7yr data alone (red shaded areas) and WMAP7yr+OHD+$H_{0}$ 
data (blue shaded areas). 
\label{fig:oLCDMOmOL}}
\end{figure}

We started by considering simple generalizations of the ``minimal" $\Lambda$CDM where the additional parameter affects directly the late-time expansion history.
For a model with an additional curvature parameter $\Omega_k$ (o$\Lambda$CDM) we obtain the constraints shown in Fig.~\ref{fig:oLCDMOmOL} and Tab.~\ref{tab:omk}. 
Figure~\ref{fig:oLCDMOmOL} shows the 1 and 2 sigmas joint constraints in the $\Omega_m$ and $\Omega_{\rm DE}$ plane; all other cosmological parameters are 
marginalized over. Table~\ref{tab:omk} shows constraints on $\Omega_m$ and $\Omega_{\rm DE}$ marginalized over all other cosmological parameters. For 
comparison we show both CMB-only constraints from WMAP7yr data and constraints for WMAP7yr combined with OHD and $H_0$ measurements; to study the 
relative improvements to the constraints due to the contribution of the Hubble constant and of the Hubble parameter alone, we also show in the table the effect of 
combining WMAP7yr+$H_{0}$ and WMAP7yr+OHD: clearly the direct constraint on the expansion history provided by OHD breaks the CMB degeneracy shrinking the 
errors on $\Omega_m$, $\Omega_{\rm DE}$, and $\Omega_{k}$ by about an order of magnitude. Table \ref{tab:omk} shows that the inclusion of the Hubble constant in the 
analysis (which alone has the same power of OHD in constraining these cosmological parameters) used as a pivot for OHD, is useful to set the final mean values of the parameters.

\begin{table}[b!]
\begin{center}
\begin{tabular}{cccc}
\multicolumn{4}{c}{\small MARGINALIZED 1D CONSTRAINTS}\\
\multicolumn{4}{c}{o$\Lambda$CDM model}\\
\hline \hline
& $\Omega_{\rm m}$ & $\Omega_{\rm DE}$ & $\Omega_{k}$\\
\hline
WMAP7yr & $0.40\pm0.19$ & $0.63\pm0.18$ & $-0.0008\pm0.068$\\
WMAP7yr+OHD & $0.280\pm0.032$ & $0.725\pm0.031$ & $-0.0039\pm0.0076$\\
WMAP7yr+$H_{0}$ & $0.245\pm0.030$ & $0.752\pm0.024$ & $0.0020\pm0.0072$\\
WMAP7yr+OHD+$H_{0}$ & $0.261\pm0.026$ & $0.743\pm0.020$ & $0.0011\pm0.0067$\\
\hline \hline
\end{tabular}
\end{center}
\caption{Constraints on $\Omega_{\rm m}$, $\Omega_{\rm DE}$ and $\Omega_{k}$ at 1-$\sigma$ obtained for an open $\Lambda$CDM cosmology.}
\label{tab:omk}
\end{table}
 
%\begin{table}[b!]
%\begin{center}
%\begin{tabular}{cccc}
%\multicolumn{4}{c}{\small MARGINALIZED 1D CONSTRAINTS}\\
%\multicolumn{4}{c}{o$\Lambda$CDM model}\\
%\hline \hline
%& $\Omega_{\rm m}$ & $\Omega_{\rm DE}$ & $\Omega_{k}$\\
%\hline
%WMAP7yr & $0.40\pm0.19(\pm0.35)$ & $0.63\pm0.18(\pm0.35)$ & $-0.0008\pm0.068(\pm0.107)$\\
%WMAP7yr+OHD & $0.280\pm0.032(\pm0.062)$ & $0.725\pm0.031(\pm0.061)$ & $-0.0039\pm0.0076(\pm0.0150)$\\
%WMAP7yr+$H_{0}$ & $0.245\pm0.030(\pm0.058)$ & $0.752\pm0.024(\pm0.047)$ & $0.0020\pm0.0072(\pm0.0142)$\\
%WMAP7yr+OHD+$H_{0}$ & $0.261\pm0.026(\pm0.051)$ & $0.743\pm0.020(\pm0.039)$ & $0.0011\pm0.0067(\pm0.0132)$\\
%\hline \hline
%\end{tabular}
%\end{center}
%\caption{Constraints on $\Omega_{\rm m}$, $\Omega_{\rm DE}$ and $\Omega_{k}$ at 1-$\sigma$ (2-$\sigma$) obtained for an open $\Lambda$CDM cosmology.}
%\label{tab:omk}
%\end{table}

\begin{figure}[t!]
\mbox{
\includegraphics[angle=0, width=0.5\textwidth]{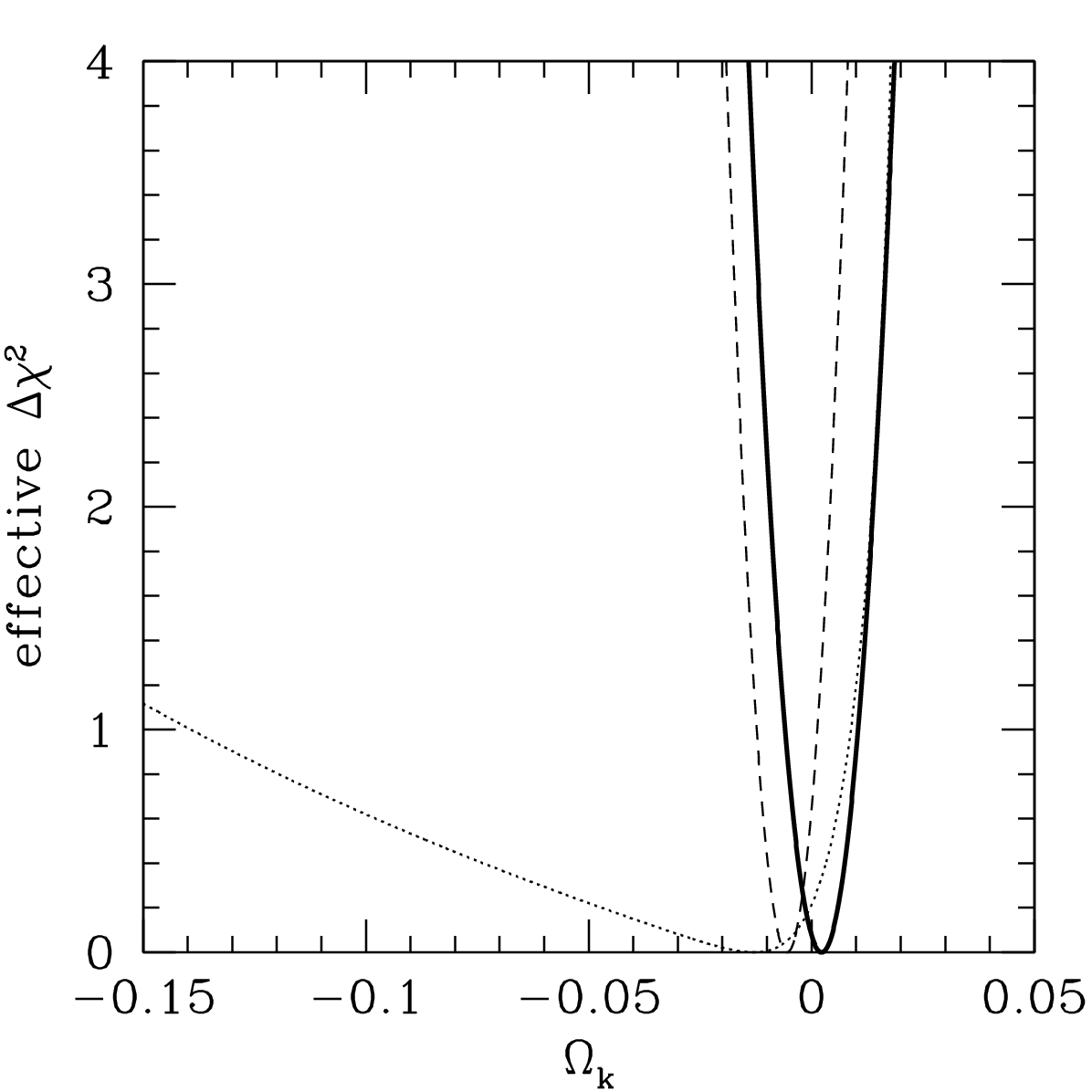}
\includegraphics[angle=0, width=0.5\textwidth]{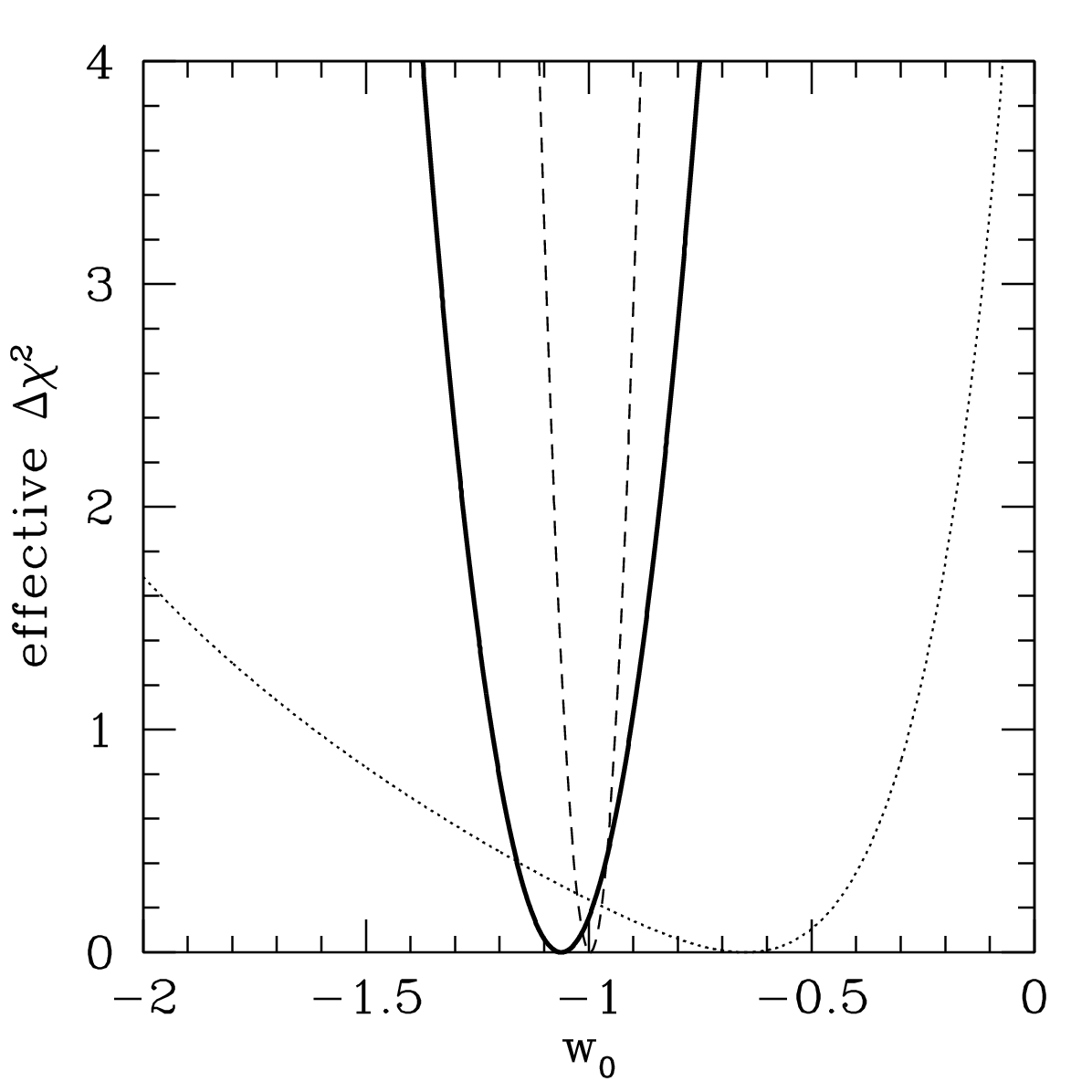}
}
\caption{Effective $\Delta\chi^{2}$ as a function of $\Omega_{k}$ (left panel) and $w_{0}$ (right panel) for an open wCDM cosmology.
The dotted lines show the constraint obtained with WMAP7yr data alone, the dashed line the constraints obtained from the combined
WMAP7yr+BAO+SNe, and the solid lines the constraints from the combined WMAP7yr+OHD+$H_{0}$. 
\label{fig:histowCDM}}
\end{figure}

We next added yet another parameter describing deviations from the standard cosmological model; we allowed the dark energy to deviate from a cosmological constant in the form of a fluid with equation of state parameter $w_{0}$ considered constant (owCDM). The resulting constraints are reported in Tab.~\ref{tab:owcdm} and in Fig.~\ref{fig:histowCDM}.
We define an effective $\Delta\chi^{2}$ as $-2 \ln P/P_{max}$, where $P$ denotes the posterior and $P_{max}$ its maximum; given a uniform prior the posterior can be identified with the likelihood. The effective $\Delta\chi^{2}$ is used only for plotting purposes.
% the confidence levels are computed integrating the posterior.

\begin{table}[b!]
\begin{center}
\begin{tabular}{ccccc}
\multicolumn{5}{c}{\small MARGINALIZED 1D CONSTRAINTS}\\
\multicolumn{5}{c}{owCDM model}\\
\hline \hline
& $\Omega_{\rm m}$ & $\Omega_{\rm DE}$ & $\Omega_{\rm k}$ & $w_{0}$\\
\hline
WMAP7yr & $0.58\pm0.31$ & $>0.52$ & $-0.013\pm0.071$ & $-0.65\pm0.61$\\
WMAP7yr+OHD & $0.277\pm0.045$ & $0.726\pm0.042$ & $0.0019\pm0.0085$ & $-0.90\pm0.18$ \\
WMAP7yr+$H_{0}$ & $0.246\pm0.021$ & $0.769\pm0.029$ & $-0.010\pm0.014$ & $-1.29\pm0.31$ \\
WMAP7yr+OHD+$H_{0}$ & $0.251\pm0.017$ & $0.751\pm0.019$ & $0.0024\pm0.008$ & $-1.06\pm0.15$\\
WMAP7yr+BAO+SNe & $0.288\pm0.017$ & $0.718\pm0.015$ & $-0.0055\pm0.0064$ &$-0.997\pm0.060$\\
\hline \hline
\end{tabular}
\end{center}
\caption{Constraints on $\Omega_{\rm m}$, $\Omega_{\rm DE}$, $\Omega_{k}$ and on the equation of state parameter $w_{0}$ at 1-$\sigma$ 
obtained for an open wCDM cosmology.}
\label{tab:owcdm}
\end{table}

%\begin{table}[b!]
%\begin{center}
%\begin{tabular}{ccccc}
%\multicolumn{5}{c}{\small MARGINALIZED 1D CONSTRAINTS}\\
%\multicolumn{5}{c}{owCDM model}\\
%\hline \hline
%& $\Omega_{\rm m}$ & $\Omega_{\rm DE}$ & $\Omega_{\rm k}$ & $w_{0}$\\
%\hline
%WMAP7yr & $0.58\pm0.31(\pm0.47)$ & $>0.52(>0.19)$ & $-0.013\pm0.071(\pm0.1078)$ & $-0.65\pm0.61(\pm1.23)$\\
%WMAP7yr+OHD & $0.277\pm0.045(\pm0.087)$ & $0.726\pm0.042(\pm0.084)$ & $0.0019\pm0.0085(\pm0.0170)$ & $-0.90\pm0.18(\pm0.45)$ \\
%WMAP7yr+$H_{0}$ & $0.246\pm0.021(\pm0.041)$ & $0.769\pm0.029(\pm0.056)$ & $-0.010\pm0.014(\pm0.028)$ & $-1.29\pm0.31(\pm0.64)$ \\
%WMAP7yr+OHD+$H_{0}$ & $0.251\pm0.017(\pm0.35)$ & $0.751\pm0.019(\pm0.037)$ & $0.0024\pm0.008(\pm0.016)$ & $-1.06\pm0.15(\pm0.29)$\\
%WMAP7yr+BAO+SNe & $0.288\pm0.017(\pm0.35)$ & $0.718\pm0.015(\pm0.030)$ & $-0.0055\pm0.0064(\pm0.0122)$ &$-0.997\pm0.06(\pm0.14)$\\
%\hline \hline
%\end{tabular}
%\end{center}
%\caption{Constraints on $\Omega_{\rm m}$, $\Omega_{\rm DE}$, $\Omega_{k}$ and on the equation of state parameter $w_{0}$ at 1-$\sigma$ (2-$\sigma$)
%obtained for an open wCDM cosmology.}
%\label{tab:owcdm}
%\end{table}

\begin{figure}[b!]
\begin{center}
\mbox{
\includegraphics[angle=0, width=0.5\textwidth]{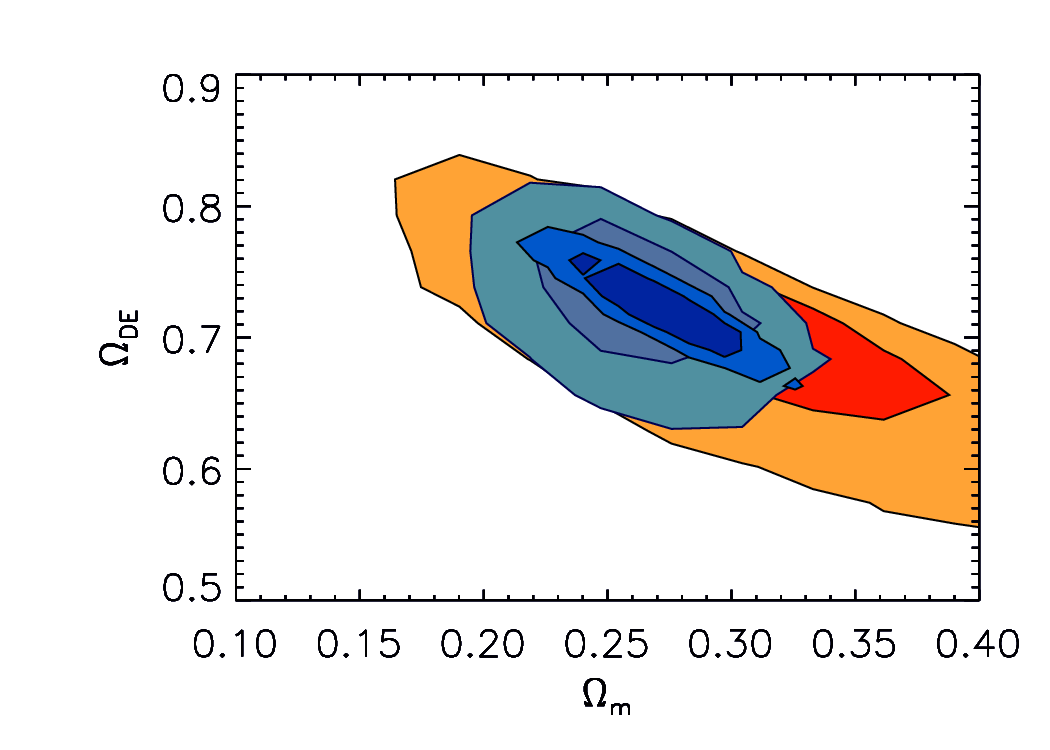}
\includegraphics[angle=0, width=0.5\textwidth]{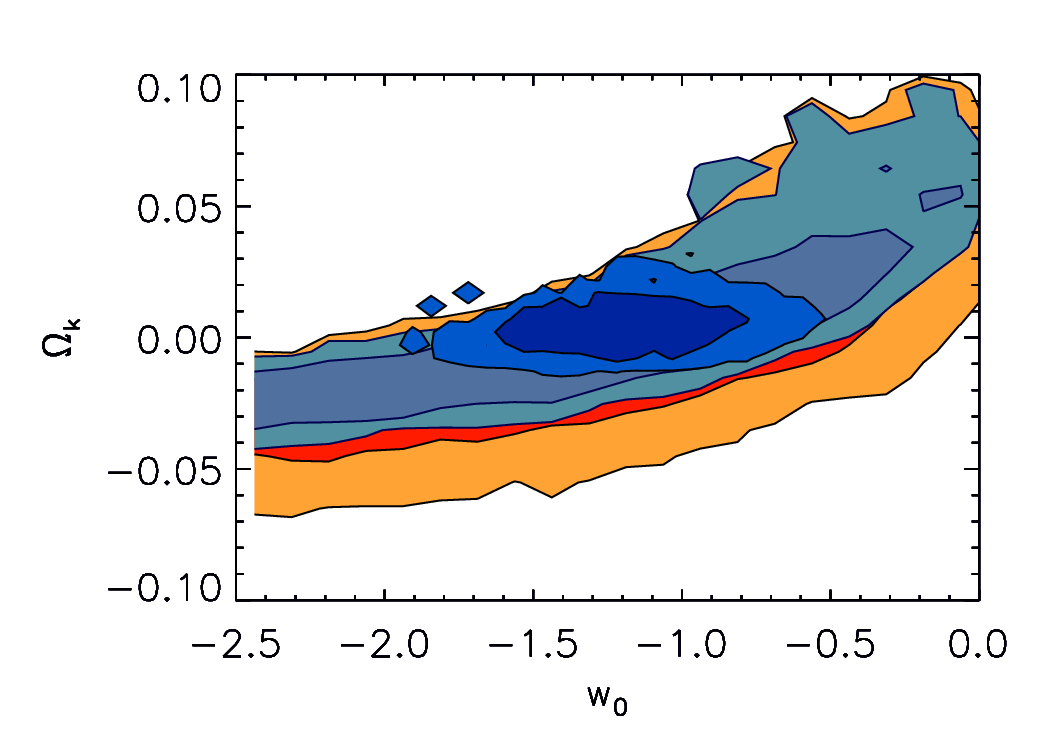}
}
\end{center}
\caption{Constraints obtained for an open-$w(z)$CDM model. Left panel: 1 and 2 sigma joint constraints in the 
$\Omega_{m}-\Omega_{\mathrm{DE}}$ plane obtained with WMAP5yr+$H_{0}$~\cite{Freedman2001} data (red shaded areas),
WMAP5yr+$H_{0}$~\cite{Riess2011} data (light-blue shaded areas), and WMAP5yr+OHD+$H_{0}$~\cite{Riess2011} data (blue 
shaded areas). Right panel: 1 and 2 sigma joint constraints in the $w_{0}-\Omega_{k}$ plane obtained with WMAP5yr+$H_{0}$~\cite{Freedman2001} 
(red shaded areas), WMAP5yr+$H_{0}$~\cite{Riess2011} data (light-blue shaded areas), and WMAP5yr+OHD+$H_{0}$~\cite{Riess2011} (blue shaded areas). 
\label{fig:ow0waCDMom}}
\end{figure}

\begin{figure}[t!]
\begin{center}
\includegraphics[angle=0, width=0.85\textwidth]{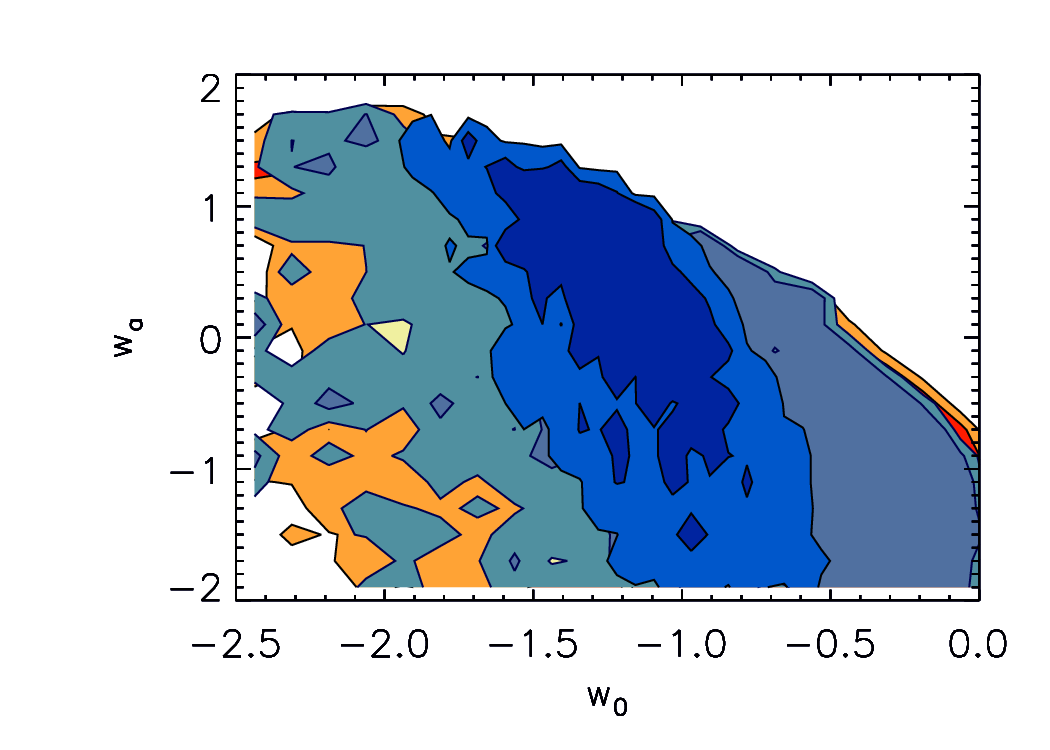}
\end{center}
\caption{1 and 2 sigma joint constraints in the $w_{0}-w_{a}$ plane
obtained for an open-$w(z)$CDM model with WMAP5yr+$H_{0}$~\cite{Freedman2001} data (red shaded areas), WMAP5yr+$H_{0}$~\cite{Riess2011} data 
(light-blue shaded areas), and WMAP5yr+OHD+$H_{0}$~\cite{Riess2011} data (blue shaded areas).
\label{fig:ow0waCDMw}}
\end{figure}

As Fig. \ref{fig:histowCDM} shows, the limits obtained on both $\Omega_{k}$ and on the dark energy parameter $w_{0}$ by WMAP7yr
data alone (dotted lines) are not very tight; if we consider the joint analysis of WMAP7yr with OHD or $H_{0}$, the limits obtained
on the cosmological parameters are significantly improved, where $H_{0}$ is best suited to constrain the adimensional density 
parameters while OHD statistical power is in constraining $\Omega_{k}$ and the dark energy EoS. When we combine WMAP7yr 
data with both $H(z)$ and $H_{0}$, we find that the constraint on $\Omega_{k}$ is improved of almost one order of magnitude, and the accuracy on the estimate of $w_{0}$
is also about three times better (see Tab. \ref{tab:owcdm} and the solid lines of Fig. \ref{fig:histowCDM}). As a comparison, we also
show how much the constraint are improved by adding to WMAP7yr data informations from BAO and SNe. We find that the combination
of WMAP7yr+OHD+$H_{0}$ has almost the same statistical power in constraining $\Omega_{k}$ as WMAP7yr+BAO+SNe. In the case of the dark 
energy parameter $w_{0}$, the WMAP7yr+BAO+SNe constraints are about 3 times better compared to the constraint obtained with the OHD; 
however we point out that the constraints on $w_{0}$ from OHD may be improved if (when) future surveys will provide an higher statistics of 
cosmic chronometers not only in the local Universe, but also at higher redshifts, where the difference in $H(z)$ due to $w_{0}$ becomes 
more significant (see Fig. \ref{fig:Hztheor}).

Finally, we further generalized our approach by exploring the constraints that can be set when not only the dark energy equation-of-state is let free
as a constant, but also allowed to vary with time, with $w(z)$ parametrized with $w_{0}$ and $w_{a}$ (o$w(z)$CDM). 
We already discussed that such a model is not already available in the LAMBDA website, and that we decided to
study the MCMC privately run from Ref.~\cite{Stern2010} (see Sect. \ref{sec:data}). Since this chain has been run with a prior on $H_{0}$ \cite{Freedman2001},
in this case we studied the effect of adding to this chain respectively the Hubble parameter measurements, the much tighter prior on $H_{0}$ provided by Ref.~\cite{Riess2011},
and both OHD and $H_{0}$. The 1 and 2 sigma joint constraints showing the $\Omega_{DE}-\Omega_{m}$, $\Omega_{k}-w_{0}$, and $w_{0}-w_{a}$ planes 
are displayed in Fig. \ref{fig:ow0waCDMom} and Fig. \ref{fig:ow0waCDMw}, and the corresponding estimates are reported in Tab. \ref{tab:ow0waCDM}.

\begin{table}[b!]
\begin{center}
\begin{tabular}{cccccc}
\multicolumn{6}{c}{\small MARGINALIZED 1D CONSTRAINTS}\\
\multicolumn{6}{c}{o$w(z)$CDM model}\\
\hline \hline
& $\Omega_{\rm m}$ & $\Omega_{\rm DE}$ & $\Omega_{\rm k}$ & $w_{0}$ & $w_{a}$\\
\hline
WMAP5yr+$H_{0}$~\cite{Freedman2001} & $0.291\pm0.064$ & $0.709\pm0.059$ & $-0.01\pm0.42$ & $-0.74\pm0.68$ & $<-0.01$ \\
WMAP5yr+OHD+$H_{0}$~\cite{Freedman2001} & $0.298\pm0.042$ & $0.697\pm0.044$ & $0.005\pm0.012$ & $-0.89\pm0.41$ & $0.56^{+0.92}_{-0.96}$ \\
WMAP5yr+$H_{0}$~\cite{Riess2011} & $0.261\pm0.025$ & $0.734\pm0.035$ & $-0.01\pm0.45$ & $-1.06\pm0.65$ & $<0.05$ \\
WMAP5yr+OHD+$H_{0}$~\cite{Riess2011} & $0.270\pm0.025$ & $0.725\pm0.026$ & $0.0040\pm0.0093$ & $-1.19\pm0.27$ & $0.56^{+0.92}_{-0.96}$ \\
\hline \hline
\end{tabular}
\end{center}
\caption{Constraints on $\Omega_{\rm m}$, $\Omega_{\rm DE}$, $\Omega_{\rm k}$, $w_{0}$ and $w_{a}$ at 1-$\sigma$ obtained for a open $\Lambda$CDM cosmology with equation-of-state parameter for dark energy parametrized as $w(z)=w_{0}+w_{a}(z/(1+z))$.}
\label{tab:ow0waCDM}
\end{table}

The red shaded areas of Fig.~\ref{fig:ow0waCDMom} and Fig.~\ref{fig:ow0waCDMw} show the constraints obtained by the original MCMC, which combined
WMAP5yr data with the $H_{0}$ estimate of Ref. \cite{Freedman2001}. Also in this case, it is evident the major improvement that is obtained by
adding to the previous dataset the $H(z)$ measurements and the $H_{0}$ estimate of Ref. \cite{Riess2011} by looking at the blue shaded areas
of the same figures. To highlight the improvement provided by the $H(z)$ measurements, we also show in light-blue shaded areas the constraints obtained 
by combining WMAP5yr data with the extremely accurate estimate of $H_{0}$ from Ref. \cite{Riess2011}.
We gain a factor three in accuracy in the constraint of $\Omega_{m}$ and $\Omega_{DE}$, a factor 40 for $\Omega_{k}$, and about a factor two for $w_{0}$.
Moreover a comparison with the constraints on $w_{0}$ and $w_{a}$ obtained by Ref. \cite{Stern2010} highlights how improved measurements 
of $H(z)$ \cite{Moresco2012} and $H_{0}$ \cite{Riess2011} help to shrink the confidence region especially around $w_{a}$. 
Including these new measurements, we set a limit of $w_{a}=0.56^{+0.92}_{-0.96}$ (68\% C.L.). It is important to stress that compared to the measurement of Ref. \cite{Stern2010},
the major contribution in the improved constraints is given by the new $H(z)$ measurements of Ref. \cite{Moresco2012}, since the same analysis
repeated with the $H_{0}$ estimate of Ref.~\cite{Riess2009} (the same used in Ref. \cite{Stern2010}), still gave compatible results with the ones discussed
above.

\subsection{Constraints on deviations from $\Lambda$CDM that affect the expansion history only through parameter degeneracies}
\label{sec:resindirect}

It has been noted already in the literature that neutrino properties can be constrained by combining CMB data with low redshift observations of the expansion rate (e.g. see Refs.~\cite{reidnu,debernardis2008,Riess2011}).

We started by considering a simple flat $\Lambda$CDM Universe where the effective number of relativistic species $N_{\rm rel}$ is let free, and not fixed to the standard value of $N_{\rm rel}=3.04$; we analyzed this model studying the combined dataset of WMAP7yr results \cite{Larson2011} with the South Pole Telescope (SPT) dataset
\cite{Keisler2011}, and of WMAP 7-years results with the Atacama Cosmology Telescope (ACT) dataset \cite{Dunkley2011,Das2011}.

\begin{figure}[b!]
\mbox{
\includegraphics[angle=0, width=0.5\textwidth]{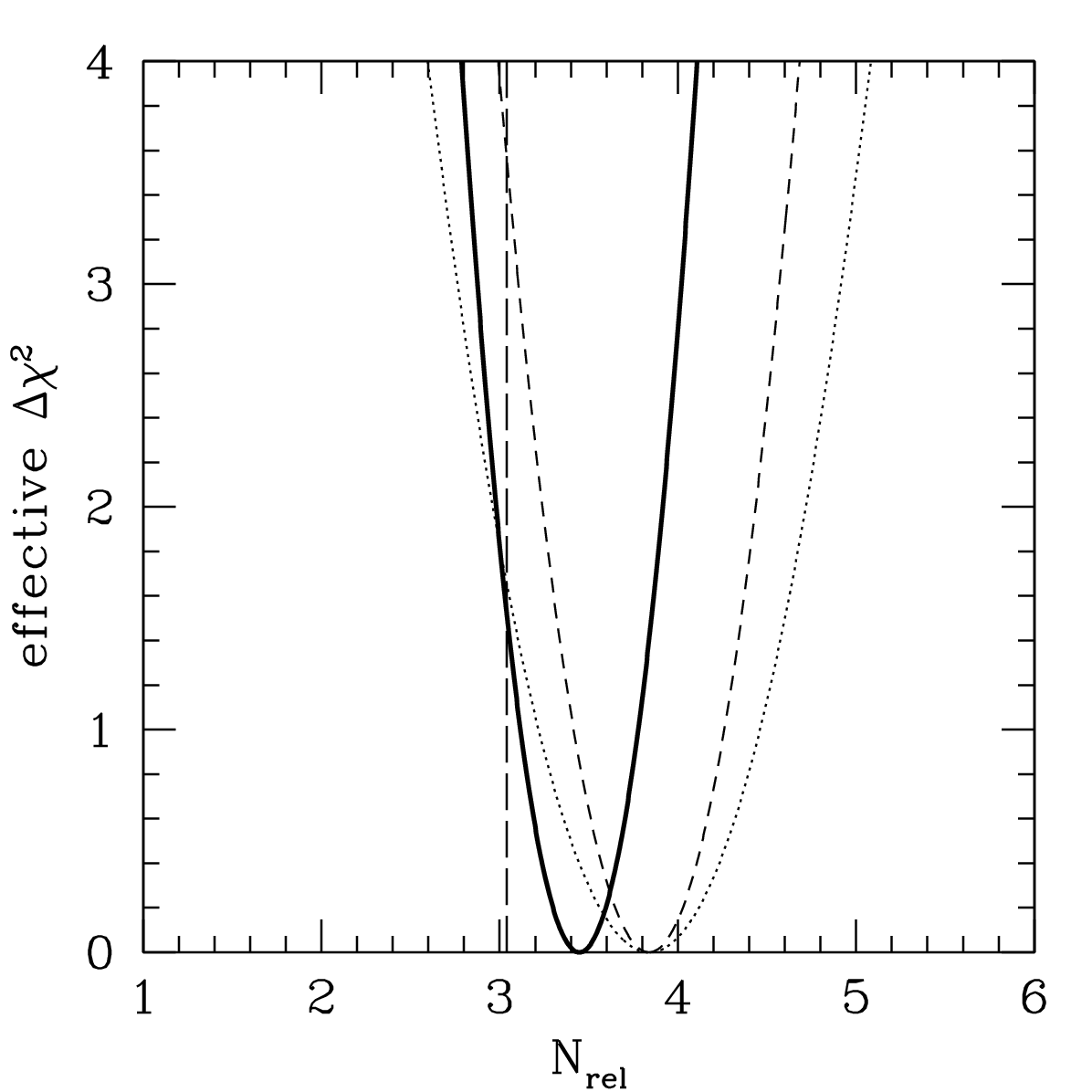}
\includegraphics[angle=0, width=0.5\textwidth]{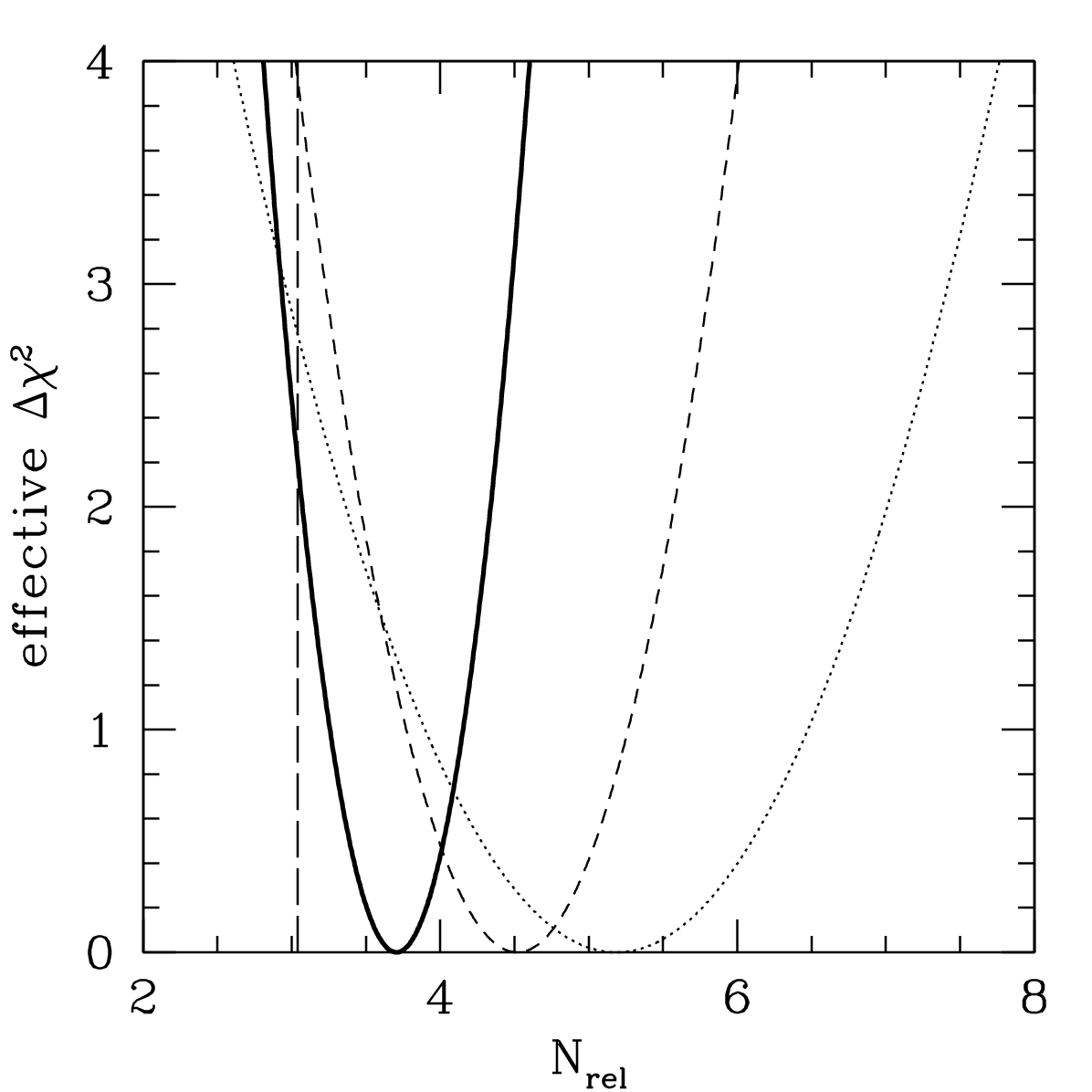}
}
\caption{Effective $\Delta\chi^{2}$ as a function of $N_{\rm rel}$. The left panel shows the constraints obtained with WMAP7yr+SPT (dotted line), 
WMAP7yr+SPT+BAO+$H_{0}$ (dashed line), and WMAP7yr+SPT+OHD+$H_{0}$ (solid line); the right panel shows the 
constraints obtained with WMAP7yr+ACT (dotted line), WMAP7yr+ACT+BAO+$H_{0}$ (dashed line), and WMAP7yr+ACT+OHD+$H_{0}$ 
(solid line). As a reference, the value $N_{\rm rel}=3.04$ is shown with a vertical dashed line. 
\label{fig:histNrel}}
\end{figure}

The effective $\Delta\chi^{2}$ as a function of $N_{\rm rel}$, for CMB data alone is shown with the dotted line of Fig. \ref{fig:histNrel}. CMB data yield a $N_{\rm rel}=3.84\pm0.62$ (at 68\% C.L.) in the case of WMAP7yr+SPT, and $N_{\rm rel}=5.19\pm1.28$ (at 68\% C.L.) in the case of WMAP7yr+ACT. 

%The constraints are better restricted when these datasets
%are combined with informations from Baryonic Acoustic Oscillations (BAO \cite{Percival2010}) and Hubble constant \cite{Riess2011},
%yielding to $N_{\rm rel}=4\pm0.5$ and $N_{\rm rel}=4.5\pm0.5$ (at 68\% C.L.), respectively for WMAP7yr+SPT+BAO+$H_{0}$ and WMAP7yr+ACT+BAO+$H_{0}$,
%which has led some authors to claim the existence of an extra relativistic specie in both cases. However this result is in conflict with what found adding to WMAP dataset the $H(z)$ measurements;
%in fact, the analysis of the combined WMAP7yr+SPT(/ACT)+OHD+$H_{0}$ not only points toward a lower value of relativistic species, with $N_{\rm rel}=3.5\pm0.5$ (at 68\% C.L.), but also
%provides the best constraints on $N_{\rm rel}$. 

These constraint are improved by adding BAO data from Ref.~\cite{Percival2010}. In fact the combinations WMAP7yr+SPT+BAO+$H_{0}$ and 
WMAP7yr+ACT+BAO+$H_{0}$ yield $N_{\rm rel}=3.84\pm0.42$ and $N_{\rm rel}=4.52\pm0.74$ (at 68\% C.L.) respectively. The fact that these measurements 
are skewed towards $N_{\rm rel}=4$ has led some authors to claim the existence of an extra relativistic specie (also called ``dark radiation'') in both cases. 
The analysis of the combination WMAP7yr+SPT(/ACT)+OHD+$H_{0}$ does not provide evidence for deviations from the standard value of three neutrinos as 
$N_{\rm rel}=3.04$ is within the 2$\sigma$ errorbar; more precisely we find $N_{\rm rel}= 3.45\pm0.33$ ($N_{\rm rel}= 3.71\pm0.45$, both at 68\%CL), and our 
data are able to exclude $N_{\rm rel}>4$ at 95\% (74\%) C.L. All the constraints on $N_{\rm rel}$ are presented in Tab. \ref{tab:Nrel}.

Of course OHD does not constrain $N_{\rm rel}$ directly but it breaks the CMB degeneracy between $N_{\rm rel}$ and the parameters setting matter-radiation 
equality and governing the expansion history (see e.g., \cite{debernardis2008}). We note how, in this case, almost all the statistical power of constraining 
$N_{\rm rel}$ is given by OHD, since the constraints obtained with WMAP7yr+SPT(/ACT)+OHD+$H_{0}$ and with WMAP7yr +SPT(/ACT)+OHD are almost
identical.

\begin{table}[h!]
\begin{center}
\begin{tabular}{ccc}
\multicolumn{3}{c}{\small MARGINALIZED 1D CONSTRAINTS}\\
\multicolumn{3}{c}{$\Lambda$CDM+N$_{\rm rel}$ model}\\
\hline \hline
& N$_{\rm rel}$ & N$_{\rm rel}$\\
& using SPT data & using ACT data\\
\hline
WMAP7yr+SPT(/ACT) & $3.84\pm0.62(\pm1.22)$ & $5.19\pm1.28(\pm2.52)$\\
WMAP7yr+SPT(/ACT)+OHD & $3.37\pm0.34(\pm0.67)$ & $3.59\pm0.48(\pm0.94)$\\
WMAP7yr+SPT(/ACT)+$H_{0}$ & $3.72\pm0.46(\pm0.89)$ & $4.37\pm0.74(\pm1.46)$\\
WMAP7yr+SPT(/ACT)+OHD+$H_{0}$ & $3.45\pm0.33(\pm0.65)$ & $3.71\pm0.45(\pm0.88)$\\
WMAP7yr+SPT(/ACT)+BAO+$H_{0}$ & $3.84\pm0.42(\pm0.83)$ &$4.52\pm0.74(\pm1.47)$\\
\hline \hline
\end{tabular}
\end{center}
\caption{Constraints on $N_{\rm rel}$ at 1-$\sigma$ (2-$\sigma$) obtained for a flat $\Lambda$CDM cosmology.}
\label{tab:Nrel}
\end{table}

\begin{figure}[t!]
\begin{center}
\mbox{
\includegraphics[angle=0, width=0.41\textwidth]{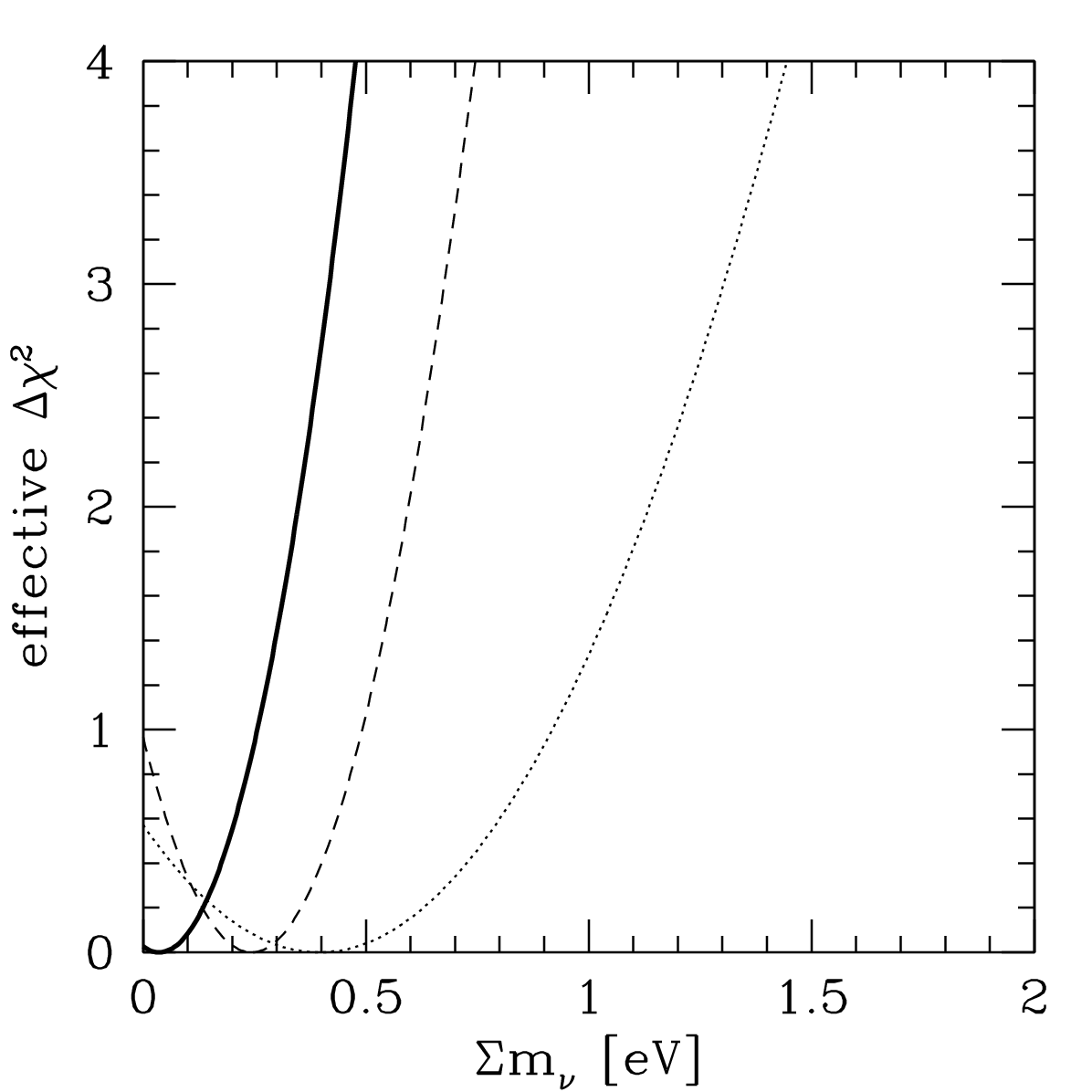}
\includegraphics[angle=0, width=0.59\textwidth]{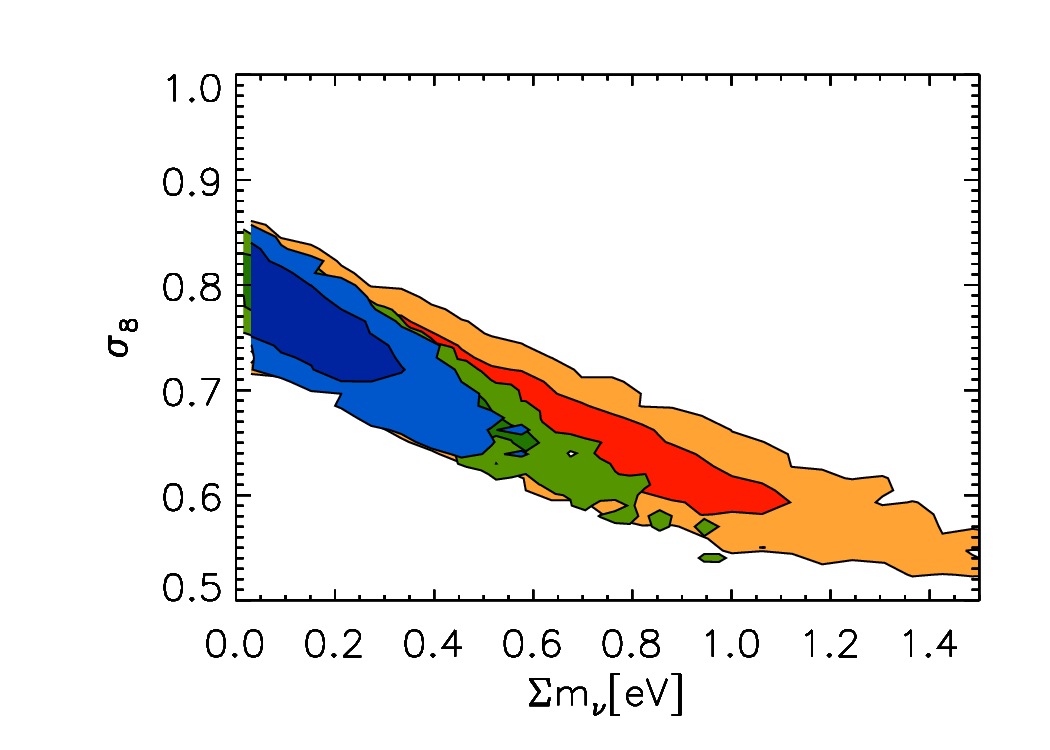}
}
\end{center}
\caption{Left panel: effective $\Delta\chi^{2}$ as a function of $\Sigma m_{\nu}$ obtained with WMAP7yr alone (dotted line), 
WMAP7yr+SN+BAO (dashed line), and WMAP7yr+OHD+$H_{0}$ (solid line). Right panel: 1 and 2 sigma joint constraints in the $\Sigma m_{\nu}-\sigma_{8}$ plane
obtained with WMAP7yr alone (red shaded areas), WMAP7yr+SN+BAO (green shaded areas), and WMAP7yr+OHD+$H_{0}$ 
(blue shaded areas). 
\label{fig:Mnus8}}
\end{figure}

We then considered a flat $\Lambda$CDM Universe where the total neutrino mass is not set to be zero but 
 %
%{\bf PS I deleted "or more accurately the total mass of relativistic species," as by fixing Nrel=3.04 in this analysis these can only be neutrinos}
%
is left as a free parameter, and looked both at the constraints
on the sum of neutrino masses\footnote{We assume three degenerate neutrino species; current data have no sensitivity on the hierarchy, so this is a very good approximation.} $\Sigma m_{\nu}$ and at the contour plot $\Sigma m_{\nu}-\sigma_{8}$. Figure \ref{fig:Mnus8} shows that WMAP7yr alone
is not able to set stringent constraints on the neutrino mass %neither in the flat $\Lambda$CDM case 
(dotted line of the left plot of Fig. \ref{fig:Mnus8}),
with $\Sigma m_{\nu}<0.76$ eV (at 68\% C.L.) and presents a clear degeneracy with the $\sigma_{8}$ parameter (see the red shaded areas of the right plot of 
Fig. \ref{fig:Mnus8}). When WMAP7yr is combined with BAO and Supernovae Type Ia (SN) datasets \cite{Hicken2009}, part of the degeneracy is lifted, and there is a 
consequent improvement also in the neutrino mass constraints, with $\Sigma m_{\nu}<0.42$ eV (at 68\% C.L.) (see the dashed line and the green shaded areas
respectively of the left and right plots of Fig. \ref{fig:Mnus8}). Also in this case, more stringent constraints are obtained by combining WMAP7yr+OHD+$H_{0}$ 
(see the blue shaded area and the continuous line of the left and right plots of Fig. \ref{fig:Mnus8} respectively), with a tight constraint on the neutrino mass $\Sigma m_{\nu}<0.24$ 
eV (at 68\% C.L.). However, we note that for this particular model, with the currently achieved precision in $H(z)$, the final constraints are mostly driven by 
the $H_{0}$ measurement. In Sect. \ref{sec:future} we will discuss the effect of more precise OHD measurements in constraining cosmological parameters.

Similarly to the $N_{\rm rel}$ case, OHD do not constrain the neutrino mass directly, but break CMB degeneracies (as for example explored in Ref.~\cite{reidnu} and references therein). All the constraints on $\Sigma m_{\nu}$ and $\sigma_{8}$ are reported in Tab. \ref{tab:mnus8}.

\begin{table} [h!]
\begin{center}
\begin{tabular}{ccc}
\multicolumn{3}{c}{\small MARGINALIZED 1D CONSTRAINTS}\\
\multicolumn{3}{c}{$\Lambda$CDM+$\Sigma m_{\nu}$ model}\\
\hline \hline
& $\Sigma m_{\nu}$ & $\sigma_{8}$\\
\hline
WMAP7yr & $<0.76$ & $0.687\pm0.075$\\
WMAP7yr+OHD & $<0.40$ & $0.742\pm0.057$\\
WMAP7yr+$H_{0}$ & $<0.24$ & $0.755\pm0.045$\\
WMAP7yr+OHD+$H_{0}$ & $<0.24$ & $0.759\pm0.045$\\
WMAP7yr+BAO+SNe & $<0.42$ &$0.734\pm0.060$\\
\hline \hline
\end{tabular}
\end{center}
\caption{Constraints on $\Sigma m_{\nu}$ (in units of eV) and $\sigma_{8}$ at 1-$\sigma$ obtained for a flat $\Lambda$CDM cosmology.}
\label{tab:mnus8}
\end{table}

\newpage
\subsection{Dependence of the results on the assumed stellar population synthesis model}
\label{sec:SPSmodels}
One of the most important sources of uncertainty in the estimate of $H(z)$ is the stellar population synthesis (SPS) model assumed. 
This issue has not be extensively studied yet, due to the intrinsic difficulties in developing new models. 
Nevertheless, it has to be taken into account, because, in principle, different models may yield to
different age (or differential age) estimates, therefore biasing the measurements of the Hubble parameter.

\begin{table}[b!]
\begin{center}
\begin{tabular}{cccccc}
\multicolumn{6}{c}{\small MARGINALIZED 1D CONSTRAINTS}\\
\multicolumn{6}{c}{o$w(z)$CDM model}\\
\hline \hline
& $\Omega_{\rm m}$ & $\Omega_{\rm DE}$ & $\Omega_{\rm k}$ & $w_{0}$ & $w_{a}$\\
\hline
WMAP5yr+$H_{0}$\cite{Freedman2001} & $0.291\pm0.064$ & $0.709\pm0.059$ & $-0.01\pm0.42$ & $-0.74\pm0.68$ & $<-0.01$ \\
WMAP5yr+$H_{0}$\cite{Riess2011}+OHD(BC03)* & $0.269\pm0.025$ & $0.729\pm0.027$ & $0.0004\pm0.011$ & $-1.25\pm0.28$ & $0.85^{+1.15}_{-1.23}$ \\
WMAP5yr+$H_{0}$\cite{Riess2011}+OHD(MaStro)* & $0.257\pm0.023$ & $0.740\pm0.026$ & $0.0011\pm0.0094$ & $-1.01\pm0.27$ & $<-0.23$ \\
\hline \hline
\end{tabular}
\end{center}
\caption{Comparison of the constraints on $\Omega_{\rm m}$, $\Omega_{\rm DE}$, $\Omega_{\rm k}$, $w_{0}$ and $w_{a}$ obtained for a open $\Lambda$CDM
cosmology with equation-of-state parameter for dark energy parametrized as $w(z)=w_{0}+w_{a}(z/(1+z))$ assuming different SPS models. *For this comparison, the $H(z)$ measurements are taken from just Ref.~\cite{Moresco2012}.}
\label{tab:ow0waCDMcomp}
\end{table}

As already discussed, to date most of the $H(z)$ estimates have been obtained using BC03 models \cite{Bruzual2003}, which are at present time the most widely used.
In Ref.~\cite{Stern2010} an effort has been done in investigating the dependence of the age estimate on the assumed SPS model, by building a mock catalog with one library and trying to recover the age with another one; in their analysis they found a general agreement, but also noted that the wavelength coverage is fundamental, since the agreement worsen noticeably in the case that the spectra analyzed do not include light bluewards of the 4000 ~\AA~ break. 
In Ref.~\cite{Moresco2011} the differential evolution of the 4000 ~\AA~ break feature as a function of redshift has been studied in a sample of ETGs extracted from SDSS to provide estimates of the Hubble constant and of the dark energy parameter $w_{0}$. These estimates have been obtained either with BC03 models and the new MaStro models, leading to the conclusion that there is no significant difference between BC03 and MaStro models, since the values for both $H_{0}$ and $w_{0}$ are compatible within the 1$\sigma$ errors. At the present time, only Ref.~\cite{Moresco2012} provides measurements of $H(z)$ using the ``cosmic chronometers'' approach with two different models; in their work they found that, with their approach, the Hubble parameter estimates have proven to be extremely robust even changing SPS model, with values always in agreement (except for one estimate at high redshift) well within the 1$\sigma$ errors.

Here, we want to study the effect on the cosmological parameters estimate of the difference in $H(z)$ due to the SPS model assumed. We, therefore, decided to compare the constraints obtained with OHD obtained with BC03 models (see Tab. \ref{tab:HzBC03}) with the ones obtained using OHD which assume MaStro SPS models; in order to perform a meaningful comparison, here we use as $H(z)$ measurements only those obtained from Ref.~\cite{Moresco2012}, not adding data from Ref.~\cite{Simon2005} and Ref.~\cite{Stern2010} which have been obtained with BC03 models. We show here as an illustrative example the case of the o$w(z)$CDM model, which is the most general. The results are reported in Tab. \ref{tab:ow0waCDMcomp}, where as a comparison are provided also the constraints obtained with just WMAP 5years data in addition to $H_{0}$ estimate from Ref. \cite{Freedman2001}.

We find the estimates of the adimensional density parameters $\Omega_{\rm m}$, $\Omega_{\rm DE}$, and $\Omega_{\rm k}$ to be extremely robust against the change of SPS model, testifying the strength and the stability of OHD in constraining the curvature and matter content in the Universe. The estimates of $w_{0}$ and $w_{a}$ are also fully compatible within the 1$\sigma$ errors, although the errors on $w_{a}$ are still large.

We therefore conclude that, given the current level of statistical errors, the analysis performed with our method is reliable, and robust against a possible 
bias to to the different choice of SPS model.

\subsection{The effect of the progenitor bias on the cosmological parameters} 
\label{sec:progbias}
The ``cosmic chronometers'' approach is based on the assumption that the chronometers that are used to probe the relative age evolution of the Universe 
are well synchronized, i.e. they have all formed at the same redshift. The progenitor bias is an effect which may falsify this assumption, altering the correct 
age-redshift relation by flattening it, since it adds to the sample an increasing younger population with decreasing redshift. It is straightforward to understand 
that the resulting Hubble parameter obtained from this flattened relation result biased toward higher values. This effect, if accurately estimated, may be corrected 
to recover the unbiased $H(z)$.

However, there are a few caveat before naively correcting for this effect. The first (and most important) one, is that there is no evidence
that our analysis is biased in that sense. As discussed before, part of our sample \citep{Simon2005,Stern2010} is based on estimates obtained from the upper-envelope
of the age-z relation, and therefore, by definition, is not affected by progenitor bias. Moreover, in Ref.~\cite{Moresco2012} the upper-envelope of the $D4000-z$ relation 
has also been analyzed, as a check, and the $H(z)$ obtained
(which should not be biased by the progenitor bias effect) is in good agreement with the Hubble parameter estimated from the median relations, showing
that the data selected in this way are still well compatible with a null contribution of the progenitor bias. 
This mainly reflects the fact that the progenitor bias is supposed to be significant especially for less massive ETGs, in which
the star formation, even if limited, is still ongoing; on the contrary, the sample analyzed here has been carefully selected to keep only the extremely
red, passive and massive ETGs exactly to avoid this issue. The other point is that the correction to be applied is model-(and also cosmology-)dependent, and 
until a more detailed knowledge of the mechanisms of ETGs formation and evolution is achieved it is difficult (if not impossible) to estimate accurately this effect, 
and then to correct for it.

Nevertheless, we decided to estimate approximately the maximum possible magnitude of this effect assuming a conservative contribution due to progenitor bias, and its impact on the 
total error of cosmological parameters. In this sense, we warn the readers that this estimate represents an upper limit to the effect of the progenitor bias,
since as discussed above our data are well compatible with being unbiased.

In Ref.~\cite{Moresco2012} is provided the theoretical formalism to correctly estimate the effect of this bias on the 
Hubble parameter, and a rough estimate is given on the basis of the actual observational constraints. On average it has been found that it is roughly half the
statistical error ($\sim0.6\sigma_{\rm stat}$), but adding only $\sim1\%$ on average to the total error budget, if also this contribution is summed in quadrature. 
Since, for the purpose of this paper, we are just interested in the effect that the progenitor bias has on the estimated cosmological parameters, we proceeded as follows: 
we considered two possible $H(z)$ relations, one which is assumed unbiased and one which we corrected taking into account the progenitor bias effect as 
estimated in \cite{Moresco2012}. The difference between the values estimated from the two relations will quantify the systematic error due to the progenitor bias.
We compared the original total error with a new total error obtained by summing in quadrature also the progenitor bias contribution, and estimated its impact by evaluating for 
each cosmological parameter $A$ the percentage difference between the old and the new error, $(\sigma_{\rm A,prog\;bias}-\sigma_{\rm A})/{\rm|A|}$.

We found that the impact of considering the progenitor bias in the total errorbars is, for each model analyzed, less than 2\% for $\Omega_{m}$, and 
less than 1\% for $\Omega_{\rm DE}$; the effect is higher for $\Omega_{k}$ simply because this value is compatible with being null. Concerning the dark energy EoS parameters
the impact is higher, but still always less than 10\% for $w_{0}$ ($\sim5\%$ when considering also the prior from the Hubble constant of Ref. \cite{Riess2011}). Concerning
the constraints on $N_{\rm rel}$, the impact is always less than 4.5\%, and for $\sigma_{8}$ less than 1\%. Finally the limit given to the sum of the neutrino masses is worsened of
about 20\% (40\% if not considering the prior on $H_{0}$).

\begin{figure}[t!]
\begin{center}
\mbox{
\includegraphics[angle=0, width=0.48\textwidth]{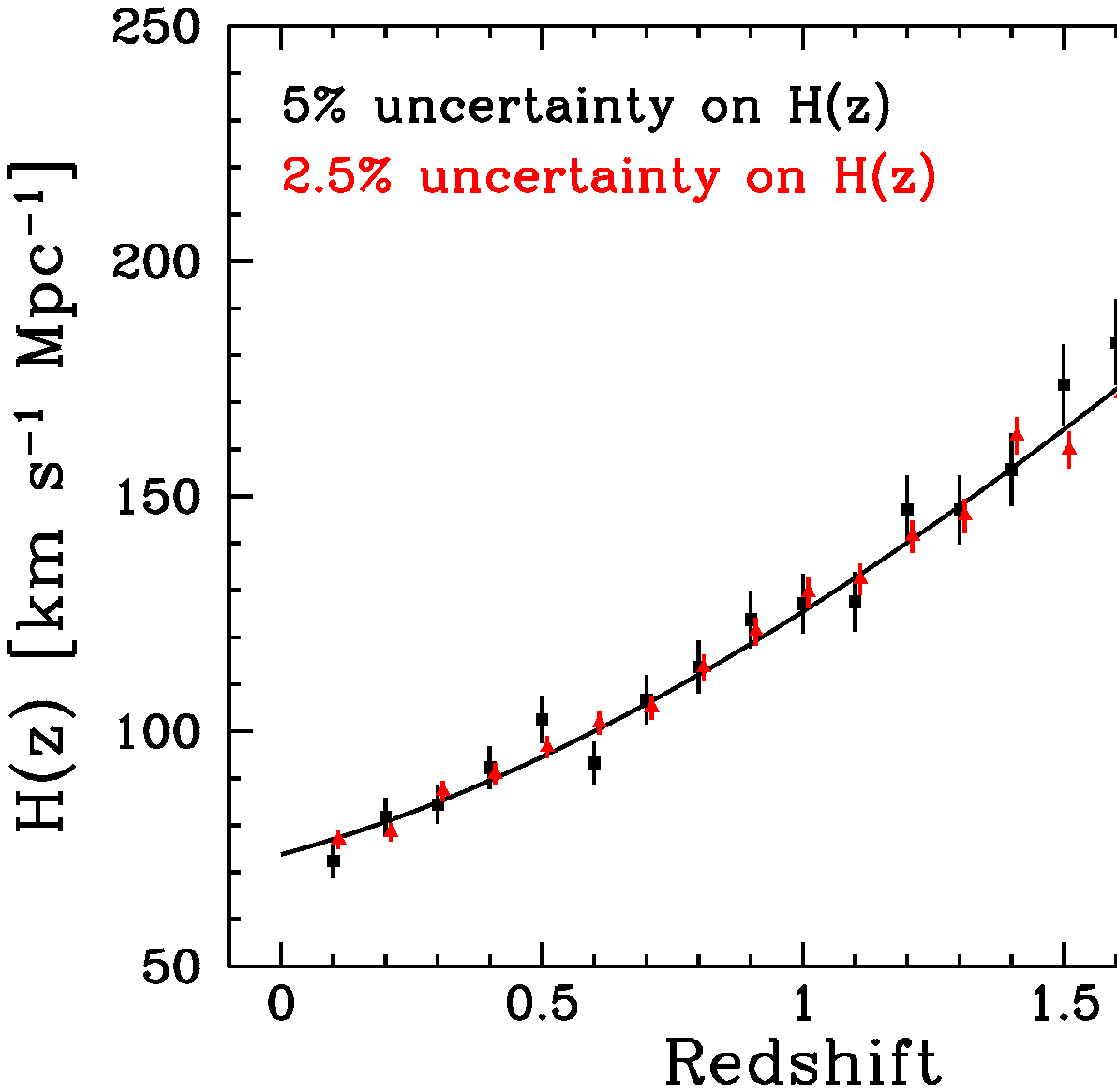}
\includegraphics[angle=0, width=0.5\textwidth]{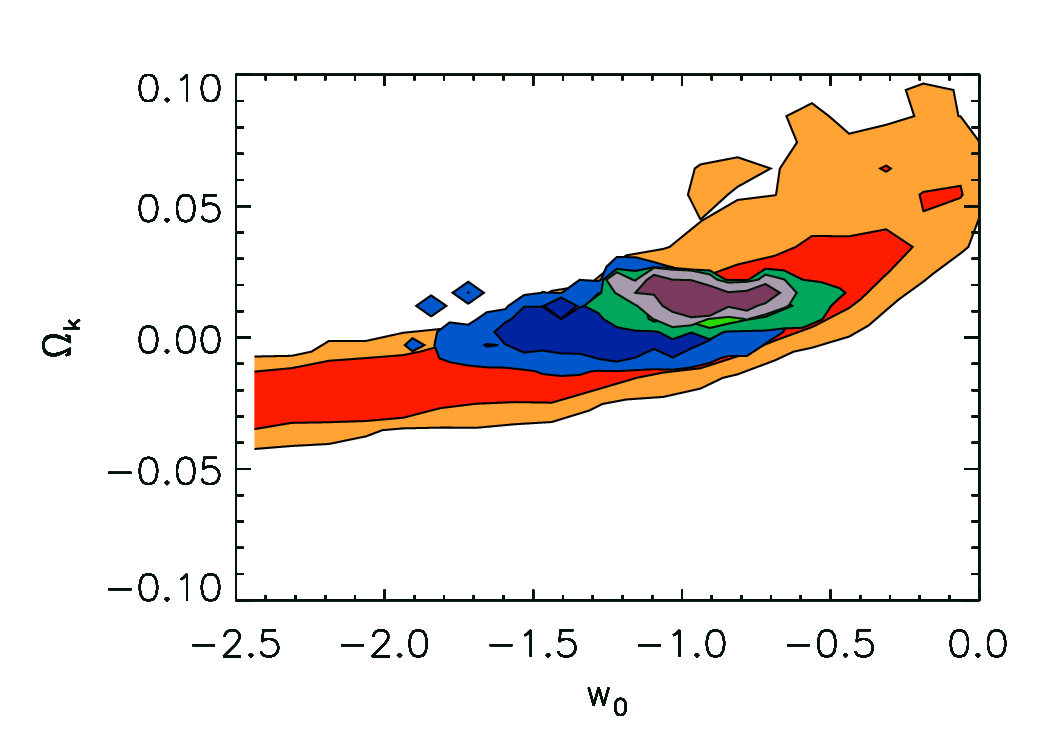}
}
\mbox{
\includegraphics[angle=0, width=0.5\textwidth]{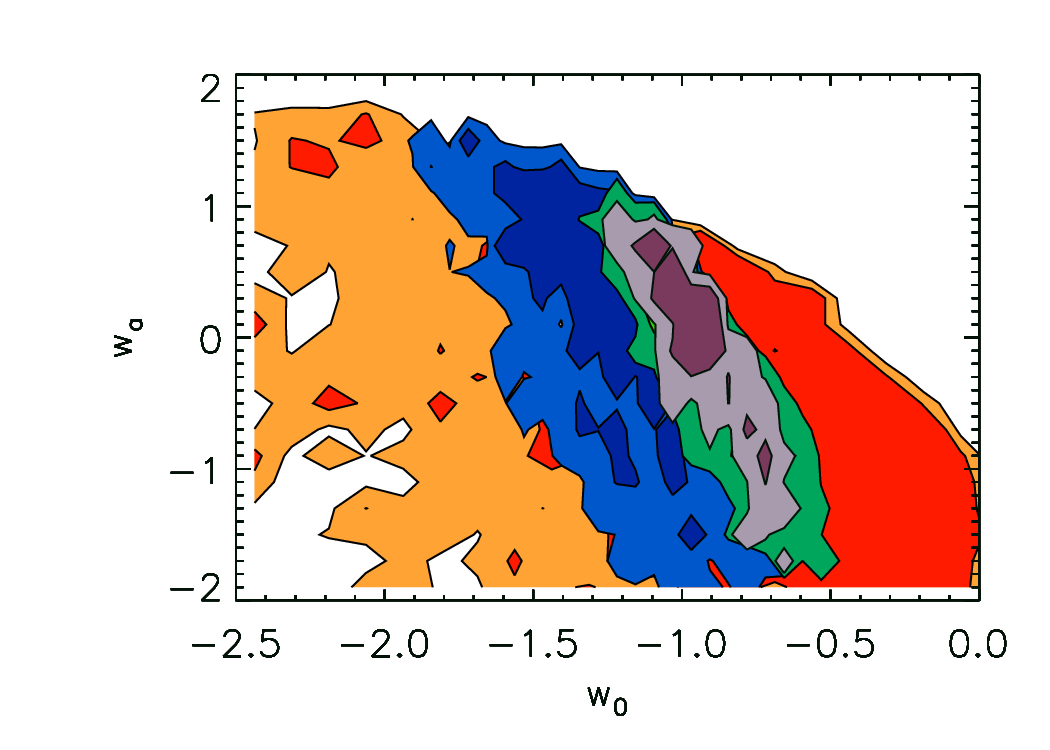}
\includegraphics[angle=0, width=0.5\textwidth]{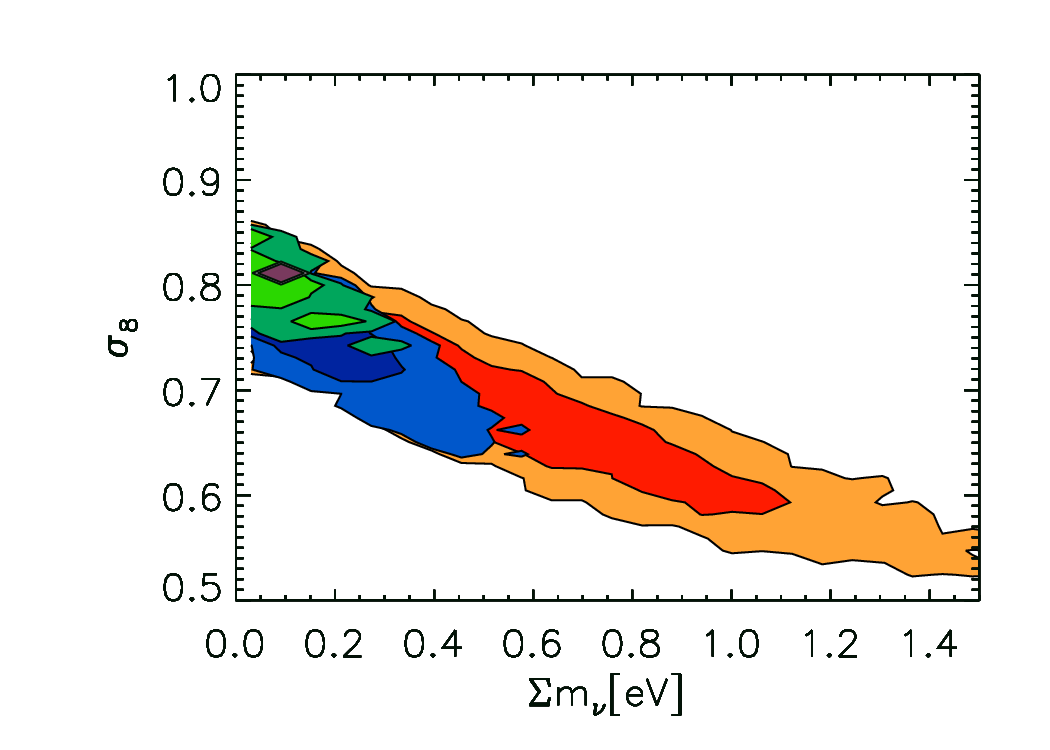}
}
\end{center}
\caption{Simulated $H(z)$ with 5\% and 2.5\% uncertainty, and relative constraints on various cosmological parameters. The red contours are obtained from WMAP data 
alone (WMAP+$H_{0}$~\cite{Freedman2001} in the case of the $w_{0}-w_{a}$ and $w_{0}-\Omega_{k}$ plots), the blue contours from the analysis of 
WMAP+OHD+$H_{0}$~\cite{Riess2011} using the observed $H(z)$ measurements, the light-green contours using the simulated $H(z)$ measurements assuming a 
5\% error on the Hubble parameter and the violet contours using the simulated $H(z)$ measurements assuming a 2.5\% error on the Hubble parameter. 
\label{fig:plotsimulHz}}
\end{figure}

\section{Future perspectives}
\label{sec:future}
At the present day, the studied OHD represent the most precise and with the widest redshift coverage $H(z)$ measurements available.
However, it has to be stressed that, unlike all the other kind of cosmological probes which nowadays can be based on dedicated surveys (e.g. BOSS \cite{Eisenstein2011} for the BAO)
these results have been obtained mainly as by-product of spectroscopic surveys devoted to other kind of studies. In this section we therefore decided to explore the possible 
improvement to the constraints of cosmological parameters if (when) a much higher statistic over a wider redshift coverage will be achieved; note that this study represents
more than a theoretical exercise, since Euclid \cite{Laureijs2011} and BOSS \cite{Eisenstein2011} will provide thousands of passive galaxies at $z>0.5$, and this will help 
to improve significantly the accuracy in $H(z)$. With the increasing statistic, it will become crucial to understand if the systematic effects may be controlled and minimized.
In Ref. \cite{Moresco2012} it is shown that the systematic error due to metallicity and star formation uncertainty can be reduced at $<1\%$ for a sample of ETGs large enough.
Moving at higher redshift, hence moving closer to the redshift of formation of these galaxies, the study of the effective SFH and of the progenitor bias will be more important,
even if some methods have already been suggested to minimize this issue (e.g. see Ref. \cite{Stern2010,Moresco2012}).

We simulated two measurements of $H(z)$ at $0.1<z<2$ with an accuracy respectively of 5\% and 2.5\%, which are shown in the upper-left panel of Fig. \ref{fig:plotsimulHz}. 
These data have been simulated assuming a flat $\Lambda$CDM cosmology, with $H_{0}=73.8{\rm \;km s^{-1}Mpc^{-1}}$ \cite{Riess2011}. The remaining three plots of Fig. 
\ref{fig:plotsimulHz} show the enhancement of the constraints that can be achieved with such measurements: the red contours represent the case of WMAP alone 
(WMAP+$H_{0}$~\cite{Freedman2001} for the $w_{0}-w_{a}$ and $w_{0}-\Omega_{k}$ plots), the blue contours the case of WMAP+OHD+$H_{0}$~\cite{Riess2011} with the $H(z)$
measurements used in this analysis, the light-green contours show how the constraints are shrunk with a 5\% error on $H(z)$ up to $z\sim2$, and the violet contours with a 2.5\%
error.

Compared to the actual measures, with a 5\% error it will be possible to reduce the errorbar on $\Omega_{k}$ of about 50{\%}, on $w_{0}$ of about 30{\%}, and on $w_{a}$ 
of about 15{\%}; similarly, with a 2.5\% error the errorbar on $w_{0}$ will be reduced of about 65{\%}, and on $w_{a}$ of about 60{\%}. The constraints on $\Sigma m_{\nu}$ 
will be even more stringent: with a 5\%-2.5\% error on $H(z)$ up to $z\sim2$ it will be possible to reduce them of about 55{\%} and 60{\%} respectively.

This forecast has been performed assuming that the errors on $H(z)$ are uncorrelated, which is the assumption that has been made in all analyses so far. 
As the statistical errors on the galaxy age determinations shrink, systematic errors become important and, in principle, they might induce correlations among $H(z)$
estimates at different redshifts. However, preliminary analyses performed so far (see sect. 3 of Ref. \cite{Stern2010}) do not find any evidence for this effect for 
spectra with sufficiently high spectral coverage and signal to noise. This is, however, an issue to consider and quantify carefully when analyzing future data.
Similarly all tests reported in Sect. \ref{sec:SPSmodels} and \ref{sec:progbias} indicate that the listed systematics are under control at the level of current data. 
A similar analysis will have to be performed on future data and of course it remains to be seen wether systematics errors can be kept under control so that 
our estimated forecasted $H(z)$ errors can indeed be achieved.

The estimate of the Hubble parameter using the ``cosmic chronometers'' appears to be a promising avenue to be further explored and investigated as a cosmological probe
complementary to the standard methods.

\section{Conclusions}
We have presented a compilation of direct measurements of the Hubble parameter as a function of redshift which includes the latest determinations and which has the 
widest redshift coverage ($0.1<z<1.75$) to date. The $H(z)$ measurements are based on the use massive and passively evolving early-type galaxies as ``cosmic chronometers'',
providing standard(-izable) clocks in the universe to trace its late-time expansion history. These measurements have been referred to as observational Hubble determination, OHD. 
The $z=0$ measurement is provided by Ref. \cite{Riess2011} using a different technique. 
This compilation has then be used in combination with CMB data to constrain cosmological parameters. As expected, the OHD are useful to break CMB cosmological parameters
degeneracies only for models that deviate from the standard (``minimal") flat $\Lambda$CDM model.
 
The cosmological parameters that most directly affect the late-time expansion history are $\Omega_m$ and $\Omega_k$ (and thus $\Omega_{\rm DE}$). If dark energy is not assumed
to be a cosmological constant but a fluid with equation of state parameter either constant in redshift or with a linear dependence on the scale factor, the equation of state parameters
also affect the expansion history directly and can thus directly be constrained by the OHD.

We reported constraints on deviations from a simple ``minimal" flat $\Lambda$CDM model that allow for spatial curvature and models that allow for spatial curvature and a dark energy
parameterized by its equation of state parameter. We found that OHD are crucial in breaking CMB parameter degeneracies and the resulting constraints are competitive (and in broad
agreement) with those obtained from BAO and/or Supernovae data.

We have also considered extensions of the ``minimal" flat $\Lambda$CDM model where the extra parameters do not affect the expansion history directly but which are constrained by
combining CMB with OHD through reduction of degeneracies with other parameters. 
We obtain competitive errors on the effective number of relativistic species, $N_{\rm rel}=3.45\pm 0.33$ at 1-$\sigma$ using as external dataset also measurements from SPT, and 
$N_{\rm rel}=3.71\pm 0.45$ at 1-$\sigma$ using as external dataset also measurements from ACT; our analysis did not find evidence for extra neutrino species, 
excluding $N_{\rm rel}>4$ at 95\% C.L. using SPT data (74\% using ACT data).
We also found very competitive limits on the sum of neutrino masses $\Sigma m_{\nu}<0.24$ eV at 1-$\sigma$.

We studied separately the combination of CMB data with OHD and $H_{0}$, to establish the relative contribution to the constraints given by these two measurements: we find that while
for many models their statistical power in constraining cosmological parameters is comparable, the $H(z)$ is particularly fundamental to constrain models in which a time-variability
of the parameters (e.g. dark energy EoS) is allowed. On the other hand, at the state of the art of the achievable precision on $H(z)$, the constraints on the sum of the neutrino mass
is almost completely given by the combination of CMB data with $H_{0}$.

The reliability and robustness of those constraints have been checked against the change of SPS model assumed to estimate $H(z)$; we find that all the cosmological parameters are
almost insensitive to the SPS model chosen, only the parameters associated to dark energy presenting a mild dependence, but still well within the 1$\sigma$ errorbars. We also
explored the effect that the progenitor bias may have on our estimates, and concluded that, even if present, given the actual statistical errors this effect should not significantly 
affect our estimate on most of the parameters.

Finally, we demonstrated that future measurements, which will provide an higher accuracy on the Hubble parameter (e.g. Euclid, BOSS), will allow to achieve an extreme accuracy for 
a variety of cosmological parameters. When measurements with an accuracy of $\sim5\%$ or lower up to $z\sim1-1.5$ will be available, the progenitor bias and the detailed star
formation history of the galaxy will become more important in the total systematic error budget, and should be carefully taken into account, and their effect quantified.

Measurements of the expansion history are the only direct indication that the Universe is undergoing an accelerated expansion and therefore that it should be dominated by a dark 
energy component. Given the magnitude of the implications of the accelerated expansion, it is important to have different, independent and complementary observations of the
expansion history. In addition to supernovae, providing the standard candles approach, and BAO, providing the standard ruler approach, the OHD yields another independent and
direct measurement of the low redshift expansion history. 
 
\acknowledgments{MM and AC acknowledge contracts ASI-Uni Bologna-Astronomy Dept. Euclid-NIS I/039/10/0, 
and PRIN MIUR ``Dark energy and cosmology with large galaxy surveys''. LV is supported by FP7-IDEAS- Phys.LSS 240117.
MM would like to thank the anonymous referee for the useful comments and suggestions, which helped to improve the paper,
and T.J. Zhang for stimulating discussions.}

\end{document}